\documentclass[review,12pt,3p,times]{elsarticle}

\usepackage{amssymb}

\usepackage{lineno}
\usepackage{setspace}
\usepackage[utf8]{inputenc}
\usepackage{amsmath}

\usepackage{booktabs}

\usepackage{bm}
\usepackage{comment}
\usepackage{float}
\usepackage{gensymb}
\usepackage{graphicx}
\usepackage{multirow}
\usepackage{nomencl}
\usepackage{natbib}
\setcitestyle{square,comma,numbers,sort&compress}
\biboptions{numbers,sort&compress}

\usepackage[version=4]{mhchem}
\usepackage{siunitx}
\usepackage{subfigure}
\usepackage{textcomp}
\usepackage{longtable,tabularx}
\usepackage{ulem}
\usepackage{verbatim}
\usepackage{xcolor}
\usepackage{soul}
\usepackage{xpatch}

\begin{document}

\begin{frontmatter}

\title{A comparative evaluation of turbulence models for simulation of unsteady cavitating flows}

\author[inst1]{Dhruv Apte \footnote{Corresponding author. Email: dhruvga@vt.edu} }
\address[inst1]{Kevin T. Crofton Department of Aerospace and Ocean Engineering, Virginia Tech, Blacksburg, VA 24060, USA}
\author[inst1,inst2]{Mingming Ge}
\address[inst2]{National Energy Technology Laboratory,Morgantown, WV 26507, USA}
\author[inst1,inst3]{Olivier Coutier-Delgosha}
\address[inst3]{Univ. Lille, CNRS, ONERA, Arts et Metiers ParisTech,Centrale Lille, UMR 9014 - LMFL - Laboratoire de Mecanique des fluides de Lille - Kampe de Feriet, F-59000 Lille, France}   

\begin{abstract}
Cavitation is a complex multiphase phenomenon characterised by vapour bubbles forming due to a sudden pressure drop and is often accompanied by increased hull vibrations, increased radiated noise and decrease in propeller and impeller performance. Although the Reynolds-Averaged Navier-Stokes (RANS) method coupled with a cavitation model is still considered a practical tool to predict cavitating flows owing to its computational efficiency, it is unable to predict the unsteadiness of vapor shedding and over-predicts the eddy viscosity. To improve the prediction, an empirical eddy viscosity correction,[Reboud et al. 1998] was proposed to consider the compressibility effects produced by cavitation. Additionally, a new type of models termed as hybrid RANS-Large Eddy Simulation (LES) models have also been recently introduced in the community, having the ability to behave as a RANS or a LES model in different regions of the flow in order to combine the computational cost efficiency of RANS with the accuracy of LES modelling. However, there exists a lack of a comprehensive review of various such turbulence models like the k-$\omega$ Shear Stress Transport Model (SST), k-$\omega$ SST Scale-adaptive Simulation (SAS), k-$\omega$ SST Detached Eddy Simulation (DES), k-$\omega$ SST Delayed Detached Eddy Simulations (DDES), Filter-Based Method (FBM) and Partially-Averaged Navier Stokes Method (PANS) to predict cavitating flows. In this work, we conduct such a review to compare their ability to predict cloud cavitating flows by comparing them with x-ray experimental data in a venturi. It is shown that with mesh refinement, standard models do show the vapor unsteadiness as seen in the experiment similar to that seen when using the Reboud correction. However, on local comparison of turbulence quantities, it is observed both forms of models have huge discrepancies with experimental data that does not improve downstream.
\end{abstract}




\end{frontmatter}

\section{Introduction} \label{introduction}
Cavitation is a complex phenomenon defined by the formation of vapour bubbles when the liquid pressure drops below the vapour pressure at a near constant liquid temperature. These bubbles travel with high speeds and collapse upon exiting the low pressure area generating shock, vibrations, erosion damage and noise that renders a detrimental effect on marine applications like propeller blades and impellers impacting their performance. The increased noise helps in detection and adversely affects submarines where survival is directly dependant on stealth. To prevent these undesired effects, it is vital to investigate the phenomenon both numerically and experimentally.

Modelling cavitating flows requires the coupling of a turbulence model and a cavitation model. Cavitation could be modelled either by separating the liquid and vapor phase by a sharp interface and reconstructing the interface using the phase volume information in each cell at each time step or using averaged methods of the equilibrium equations to understand flow parameters. Various models of systems of equations can be used in the averaged-interface approach from the seven equation model \citep{zein2010modeling} to the three equation model. These models use either different sets of equations for each phase or assume both phases are mixed well and connected by the void fraction equation. A notable mention is the Transport Equation Model (TEM) where the additional transport equation contains condensation and vaporization terms. The advantage of using TEMs is the condensation and vaporization terms can be treated separately for analysis. Notable TEM models include the ones by Merkle \citep{merkle1998computational}, Kunz \citep{kunz2000preconditioned} and Schnerr-Sauer \citep{sauer2000unsteady}.

Regarding the turbulence modelling, Direct Numerical Simulations (DNS) seems the ideal approach as it resolves all the turbulence scales however its computational cost makes it out of reach for most of cavitating flow applications. 
Large Eddy Simulations (LES) where the smaller scales below a sub-grid scale limit are modelled and rest are resolved have been conducted \citep{gnanaskandan2016large}$^,$ \citep{bensow2010implicit} but it is still computationally expensive. Hence Reynolds-Averaged Navier-Stokes (RANS) methods are the primary methods for studying the flows. 
In the recent years, a new type of turbulence models have been developed. Termed as hybrid RANS-LES models, these models behave as RANS near the wall boundary and LES away from the wall. This transition is governed either by a sharp interface  or a smooth transition towards RANS as we move towards the wall. These models are thus able to combine the computational efficiency of RANS with the accuracy of LES. 
Different modelling approaches have been used to numerically investigate cavitating flows. Reboud \textit{et al.} \citep{reboud1998two} introduced an empirical eddy viscosity limiter to suppress the turbulent eddy viscosity at the interface. Coutier-Delgosha \textit{et al.}\citep{coutier2003evaluation, coutier2003numerical} compared the k-$\epsilon$ re-normalization group (RNG) and the k-$\omega$ models with and without the Reboud correction and observed the standard models are in poor agreement with the experiments and the correction remarkably improves the model's ability to predict the vapor shedding behaviour: while the standard model predicts a quasi-steady cavity, the model with corrections predicts an unsteady vapor cavity that agreed well with the experiments. Seo and Lele \citep{seo2009numerical} implemented the correction in the Spalart-Allmaras model to study the partial and cloud cavitating flow around a NACA66 hydrofoil and concluded similar observations along with observing the instability on the shear layer upstream due to the re-entrant jet.  Zhang \textit{et al.} \citep{zhang2021compressible} conducted similar studies using the k-$\omega$ SST (Shear Stress Transport) model with and without the correction for flow inside a venturi and observed that the correction significantly improves the Reynolds shear stress data near the wall. Hidalgo \textit{et al.} \citep{hidalgo2019scale} conducted 3D calculations of unsteady cloud cavitation around a NACA66 hydrofoil using the k-$\omega$ SST Scale Adaptive Simulation(SAS) model where the model was able to produce a regular detached cavity. For further detailed readings into hybrid RANS-LES models, the reader is encouraged to refer \citep{chaouat2017state} and \citep{heinz2020review}.

In regards to the hybrid RANS-LES models, there have been studies performed to compare a hybrid model with a standard RANS model in predicting cavitating flows. Another approach, called the Filter-Based Method (FBM) was used by Wu \textit{et al.} \citep{wu2005time} and compared with the Launder and Spalding k-$\epsilon$ model (LSM) for 2D calculations of simulating cavitating flows around a Clark-Y hydrofoil with an interfacial dynamics-based cavitation model. They showed that the FBM model was able to predict the cavity breakup better and lower eddy viscosity than the LSM model. Zhang \textit{et al.} \citep{zhang2016hybrid} modified the model by replacing the k-$\epsilon$ model by the k-$\omega$ SST model as the baseline model to conduct a 2D simulation of cloud cavitating flow over a NACA66 hydrofoil and able to reproduce the cavity growth, shedding and collapse accurately. Kim \citep{kim2009numerical} compared calculations for cavitating flow over a hydrofoil by comparing the standard Wilcox k-$\epsilon$ model, the realizable k-$\epsilon$ model with the hybrid model Detached Eddy Simulations (DES) and LES calculations. In the study, both the DES and LES were able to accurately predict the oscillating sheet cavity, the formation and regular shedding of the vapor cloud cavity and the lift and drag forces. Wang \textit{et al.} conducted a similar study comparing the ability of standard RANS models with DES and LES to simulate cavitating flow around a NACA66 hydrofoil focusing on their ability to simulate the shock wave propagation process. Their results demonstrated that while DES was able to well predict the cavity morphology, it could not accurately predict the high pressure generated as a result of shock wave propagation and the flow field characteristics near the wall.Bensow \textit{et al.} \citep{bensow2011simulation} compared simulating cavitating flow around a twisted hydrofoil using implicit LES, k-$\epsilon$ with the Reboud correction and Delayed Detached eddy Simulation (DDES), an improved version of the DES model. They summarized the DES calculation was similar to the LES one but with a weaker re-entrant jet formation. Unlike the standard RANS model, the DES simulation did not over-predict the turbulent eddy viscosity around the trailing edge of the cavity. Similarly, Vaz \textit{et al.} \citep{vaz2017improved} conducted calculations to evaluate the influence of the turbulence model. However, their study which compared the k-$\sqrt{k}$L model as the RANS model (with and without the Reboud correction) and Delayed DES model based on the k-$\omega$ SST model determined that the cavity growth and detachment were better captured with the RANS with Reboud correction model than the latter and was accounted to the boundary layer shielding from the turbulence resolution. Huang \textit{et al.} \citep{huang2017numerical} used a slight modification of another model, titled Partially-Averaged Navier-Stokes (PANS) model  where the ratio of unresolved kinetic energy to total kinetic energy ($f_k$) was varied spatially and temporally to predict sheet/cloud cavitating flow over a hydrofoil. The study was able to capture the transient evolution of the vapor cavity well using the modified and original model while the modified model was also able to reproduce the unsteady vapor shedding behaviour. Kanfoudi \textit{et al.} \citep{kanfoudi20183d} conducted a similar 3D investigation of cavitating flow over a hydrofoil using the PANS model based on the k-$\omega$ SST model achieving similar results predicting the unsteady vapor shedding and the re-entrant jet that agreed well with experimental data. A list of recent studies with the turbulence models used is shown in Table \ref{tab:table1}.  Although the above-mentioned studies are able to reproduce cavitating flows globally, they are unable to predict cavitating flow on a local scale by comparing with the local profiles of time-averaged velocity and/or Reynolds shear stress and turbulent kinetic energy. Furthermore, these studies compare 1-2 hybrid RANS-LES models to a standard RANS model and experimental/LES data. Thus, there exists a need in the community to conduct an extensive review of hybrid RANS-LES models and their ability to accurately predict unsteady cavitating flow since it will have the potential to provide an updated, or a new direction in the modelling of unsteady cavitating flows.

The recent availability of high-fidelity data from X-ray experiments \citep{ge2021cavitation} for cloud cavitating flow inside a small venturi channel provides a reliable set of turbulent quantities like the turbulent kinetic energy and Reynolds stress. This data makes it possible to assess various RANS and hybrid RANS-LES models for a universal testcase and gain insights into the turbulence model influence. The goal of this work is assess the ability of various turbulence models to predict cavitating flows by comparing it with the experimental data in both a global and local perspectives. 
The rest of the paper is as follows: In Section \ref{NM}, the numerical models detailing the cavitation model and the turbulence model equations are presented. Section \ref{testcase} discusses the test case with the numerical setup while Section \ref{results} presents the results evaluating the various turbulence models on their ability to predict global and local behaviour of cavitating flows. Section \ref{conclusion} concludes the paper. 

\section{Numerical Model}\label{NM}
\subsection{Basic Governing Equations}
In this work, we use the Transport-Equation Model approach (TEM) where the two phases (liquid and vapor) are strongly coupled. The cavitation process is governed by the momentum and mass transfer equations defined as:

\begin{eqnarray}
\frac{\partial (\rho_{m} u_{i})}{\partial t}+\frac{\partial (\rho_{m} u_{i} u_{j})}{\partial x_{j}}\nonumber\\
=-\frac{\partial p}{\partial x_{i}}+\frac{\partial}{\partial x_{j}}[(\mu_{m} +\mu_{t})(\frac{\partial u_{i}}{\partial x_{j}}+\frac{\partial u_{j}}{\partial x_{i}}- \frac{2}{3}\frac{\partial u_{k}}{\partial x_{k}} \delta_{ij})]
\end{eqnarray}

\begin{equation}
    \frac{\partial\rho_{l} \alpha_{l}}{\partial t}+\frac{\partial(\rho_{l} \alpha_{l} u_{j})}{\partial x_{j}}= \dot{m}^{-} + \dot{m}^{+}
\end{equation}
\begin{equation}
    \rho_{m}= \rho_{l}\alpha_{l}+\rho_{v}\alpha_{v}
    \label{eq:mixture_density}
\end{equation}
\begin{equation}
    \mu_{m}= \mu_{l}\alpha_{l}+\mu_{v}\alpha_{v}
\end{equation}
where $u_{j}$ is the velocity component in the jth direction, $\rho_{m}$ and $\mu_{m}$ are respectively the density and viscosity of the mixture phase, \textit{u} is the velocity, \textit{p} is the pressure, $\rho_{l}$ and $\rho_{v}$ are respectively the liquid and vapor density, $\mu_{l}$ and $\mu_{v}$ are respectively the liquid and vapor dynamic viscosity while $\mu_{t}$ represents the turbulent viscosity. $\alpha_{l}$ and $\alpha_{v}$ are respectively the liquid and vapor void fraction. The source ($\dot{m}^{+}$) and sink ($\dot{m}^{-}$) terms represent the condensation (vapor destruction) and evaporation (vapor formation) terms respectively and will be discussed below.

\subsection{Cavitation Model}
The Merkle computational model \citep{merkle1998computational} has been used throughout this work for the cavitation model. This model is derived for a cluster of bubbles, as cavitation occurs in general rather than just a single bubble. The mass fraction form of the condensation and destruction terms is defined as:
\begin{equation}
    \dot{m}^{-}= \frac{C_{dest} min(p-p_{sat},0)\gamma \rho_{l}}{0.5 U_{\infty}^{2} t_{\infty}\rho_{v}}
\end{equation}
\begin{equation}
    \dot{m}^{+}= \frac{C_{prod} max(p-p_{sat},0)(1-\gamma)}{0.5 U_{\infty}^{2} t_{\infty}}
\end{equation}
Where $\gamma$ is the liquid volume fraction, $\rho_{v}$ and $\rho_{l}$ are the vapor density and the liquid density respectively, $p$ and $p_{sat}$ are the pressure and the saturation pressure respectively, $t_{\infty}$ is the free stream time scale and $U_{\infty}$ is the free stream velocity.  The empirical factors $C_{dest}$ and $C_{prod}$ are set as 80 and 1e-3 respectively. These values are the default values used in the solver.

\subsection{Turbulence Models}

Here, a range of various RANS and hybrid RANS-LES models are implemented and investigated for the study. As mentioned in Section \ref{introduction}, hybrid RANS-LES models behave as a RANS model near the boundary wall and as a LES model away from the boundary wall thus setting up a balance between computational cost and accuracy. They are further divided into two categories base on the transition from RANS to LES and vice versa-the zonal methods where the RANS model and subgrid LES model regions are divided by a sharp interface and the models equations containing a sharp 'switch' from RANS to LES and the non-zonal methods who have a more smooth,continuous transition either using clipping parameters or a continuous formalism. We will briefly discuss both of them with the models used in this study .

\subsubsection{RANS models}
The general methodology is based on the statistical averaging of the quantities in the Navier-Stokes equations. Here, the liquid and vapor are assumed to be strongly coupled and the slip in phase interface is neglected.To ensure the coupling, the mixture density term $(\rho_{m})$ from Eq. \ref{eq:mixture_density} is implemented in the traditional RANS equations.

Originally designed for simulating steady flows, they had been increasingly used for simulating aeronautical or turbomachinery applications until recently. Although computationally economic, they are not able to predict some major phenomena of turbulence as accurately as observed in experiments. In this study, we initially use the standard k-$\epsilon$ model \citep{launder1983numerical}. The standard k-$\epsilon$ model is often not able to predict the dynamics close to the wall. To alleviate this issue, wall functions can be implemented or the model can be blended with the k-$\omega$ model \citep{wilcox1998turbulence}  where the k-$omega$ is utilized in the boundary region and k-$\epsilon$ is utilized in the free shear flow region to form the k-$\omega$ SST \citep{menter2003ten} model. In this study, we employ both procedures to evaluate their ability to predict cloud cavitation.

\subsubsection{Zonal Hybrid RANS-LES models}
Zonal hybrid models are characterised by the sharp separation interface between regions modelled by RANS and LES respectively. The sharp interface, however results in continuity issues and predicts incorrect velocity and turbulence properties profiles near the interface. Thus, an internal forcing is generally required to resolve it.  One of the leading zonal models is the Detached Eddy Simulation (DES97) model developed by Spalart \textit{et al.} \citep{spalart1997comments} where the model switch occurs as a function of the turbulent length scale. Spalart \textit{et al.} \citep{spalart2006new} concluded the original model exhibited an incorrect behaviour in the thick boundary layers when the grid spacing parallel to the wall is less than the boundary layer thickness. They proposed a delayed DES model (DDES) \citep{spalart2006new} with a RANS mode in the thick boundary layers and allowing LES after separation with an improved function for the length scale. Researchers have extended the DES to other two-equation models also after the Spalart-Allmaras model, one of the notable one being the k-$\omega$ SST-DES model developed by  Menter \textit{et al.} \citep{menter2003ten} 

\subsubsection{Non-zonal hybrid RANS-LES models}

Non-zonal models are able to bridge the two different models in a more continuous way. Several such non-zonal models are implemented and used to predict cloud cavitating flows in this study. The Partially-Averaged Navier Stokes method (PANS) developed by Girimaji \textit{et al.}  \citep{girimaji2003pans, girimaji2006partially, girimaji2006partially2} utilises the ratio of subfilter energy to total energy ($f_{k}=k_{sgs}/k$) and the ratio $f_{\epsilon}=\epsilon_{sgs}/\epsilon$ as the cutoff parameters. The ratios can either be empirically fixed throughout the calculation or varied spatially and/or temporally as a function of the turbulent length scale. If $f_{k}$ is close to zero, the simulation will be close to DNS while $f_{k}=1$ implies the simulation will be a RANS calculation. The model has been implemented using the classical RANS equations of both the k-$\epsilon$ and k-$\omega$ SST model \citep{lakshmipathy2006partially}. Another similar model is the Filter-Based Method (FBM) \citep{johansen2004filter} which uses a filtering approach to determine the eddy viscosity based on the k-$\epsilon$ model in terms of filter size and turbulent length scale. With the increase in filter size closer to the wall, the model smoothly approaches the standard k-$\epsilon$ model. Another notable non-zonal model used in this study is the Scale-adaptive Simulation (SAS) model \citep{menter2005scale, menter2010scale} which implements the Von-Karman length scale into RANS models thus making a transition to LES in unsteady flow field while being a RANS model in the steady flow field. It is thus important to note that SAS does not transition based on the grid but on the localized flow physics. 
 Seven RANS and hybrid RANS-LES models were employed for this study. Table \ref{tab:table3} lists all the models used along with their original references and previous studies conducted that used those respective models to evaluate their efficiency along with the model abbreviations that will be used throughout this study. 
\begin{table*}[htbp]
\caption{\label{tab:table3} List of the turbulence models employed in the present study}
{\begin{tabularx}{\linewidth}{lXXX} \toprule
 Turbulence Model & Original paper & Previous studies conducted & Abbreviation \\ \midrule
 \hline
Standard k-$\epsilon$ & Launder $\&$ Spalding \citep{launder1983numerical} & Kim, 2009 \citep{kim2009numerical}, Nouri \textit{et al.} \citep{kumar2020assessment}, Leclerq \textit{et al.} \citep{leclercq2017numerical} & kEpsilon \\
k-$\omega$ SST & Menter \textit{et al.} \citep{menter2003ten} & Li \textit{et al.} \citep{li2019calculation}, Zhang \textit{et al.} \citep{zhang2021compressible}, Geng $\&$ Escaler \citep{geng2020assessment}, Chebli \textit{et al.} \citep{chebli2021influence} & kOmegaSST \\
k-$\omega$ SST SAS & Menter \textit{et al.} \citep{menter2005scale, menter2010scale}  & Decaix \textit{et al.} \citep{decaix2012time} $^,$ \citep{decaix2013investigation} , Hidalgo \textit{et al.} \citep{hidalgo2019scale}, Ennouri \textit{et al.} \citep{ennouri2019numerical}  & kOmegaSST-SAS \\
k-$\omega$ SST DES & Spalart \textit{et al.} \citep{spalart1997comments} & Sedlar \textit{et al.} \citep{sedlar2016numerical}, Gao \textit{et al.} \citep{gao2012hybrid}, Wang \textit{et al.} \citep{wang2021comparative} & kOmegaSSTDES\\
k-$\omega$ SST DDES & Spalart \textit{et al.} \citep{spalart2006new} & Bensow \textit{et al.} \citep{bensow2011simulation} , Vaz \textit{et al.} \citep{vaz2017improved}, Wang \textit{et al.}. \citep{hwang2021numerical} & kOmegaSSTDDES  \\
Filter-Based Method (FBM) & Johansen \textit{et al.} \citep{johansen2004filter} & Zhang \textit{et al.} \citep{zhang2016hybrid}, Liu \textit{et al.} \citep{liu2021numerical}, Sun \textit{et al.} \citep{sun2019numerical} & FBM \\
PANS (based on the k-$\omega$ SST model) & Lakshmipathy and Girimaji \citep{lakshmipathy2006partially} & Huang \textit{et al.} \citep{huang2017numerical}, Kanfoudi \textit{et al.} \citep{kanfoudi20183d}, Ji \textit{et al.} \citep{ji2013numerical}, Zhou \textit{et al.} \citep{zhou2019development} & PANS \\
\end{tabularx}}
\end{table*}

\subsubsection{Reboud correction}

Cloud cavitating flows are an extremely unsteady phenomenon with fluctuations at various stages. It is characterized by periodic shedding of the primary cavity from the wall followed by a secondary detachment downstream , eventually breaking up in the zone of high pressure. Conventional RANS models have been unable to reproduce the vapor shedding as they over predict the eddy viscosity in the cavitating region and thus prevent the formation of re-entrant jet that induces the cloud separation and is thus, necessary for the periodic shedding. To capture, the periodic shedding, Reboud \textit{et al.} \citep{reboud1998two} introduced an empirical eddy viscosity limiter modification that reduces the eddy viscosity significantly in the cavitation region based on the vapor void fraction. The modification is defined as:
\begin{equation}
    f(\rho_{m})=\rho_{v} + (1-\beta)^{n} (\rho_{l}-\rho_{v})
\end{equation}
where~$\rho_{m}$ is the mixture density,~$\beta$ is the vapor void fraction,~$\rho_l$ and~$\rho_v$ is the density of liquid and vapor, respectively. If the void fraction is zero, $f(\rho)$ will be equal to liquid density in the model equation. If the void fraction increases to 1, $f(\rho)$ will be the vapor density in the model equation. The value of $n$ is empirical and various studies have used different values. \citep{coutier2003numerical, goncalves2011numerical, zhang2021compressible, chebli2021influence}. In this study, $n$ is taken as 20. With this modification, it has been observed that the model is able to predict an unsteady cavity that agree with the experiments.

\section{Test Case}\label{testcase}
\subsection{Simulation set-up}
\begin{figure*}
\includegraphics[width=15cm, height=3cm]{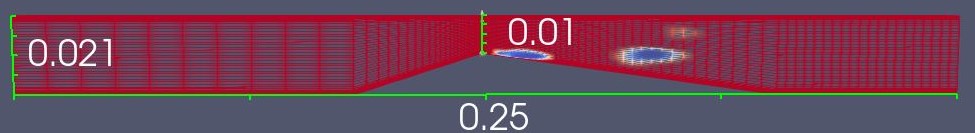}
\caption{The venturi-type geometry}
\label{venturi}
\end{figure*}

\begin{figure}
\includegraphics[width=8cm, height=4cm]{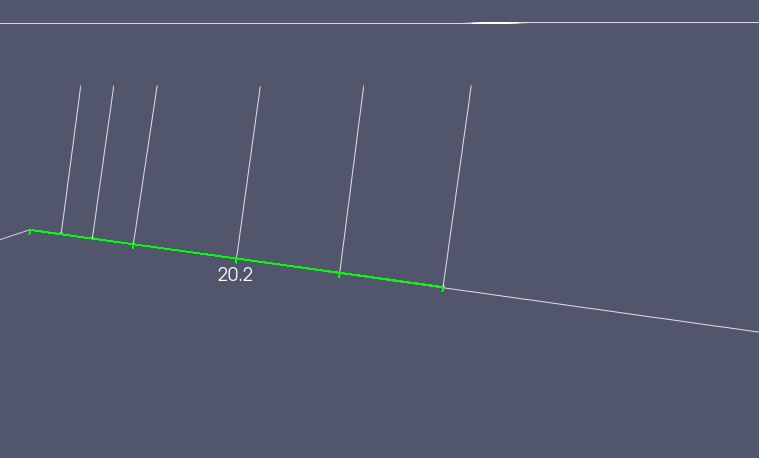}
\caption{Profiles used for local comparisons. (dimensions in mm)}
\label{profiles}
\end{figure}

X-ray high-speed Particle-Image Velocimetry experiments were conducted to obtain a reliable set of data for cloud cavitating flows including velocity fluctuation data. For details about the experimental setup, readers are referred to Ge \textit{et al.} \citep{ge2021cavitation}. A venturi-type geometry is used in this work, shown in Fig \ref{venturi} having an 18 \degree convergent angle and an 8 \degree divergent angle. The height of the venturi channel is originally 21 mm and reduces to 10 mm at the throat. For consistency with the experiments, the inlet velocity is set to 8.38 m/s. A specific characteristic defining a cavitating flow is the cavitation number ($\sigma$) expressed as the ratio of pressure force to inertial force. The cavitation number indicates whether cavitation has occurred or not and is expressed as:
\begin{equation}
    \sigma = \frac{2(p-p_{v})}{\rho_{l}u^{2}}
    \label{eq:cav_number}
\end{equation}
where $p_{v}$ is the vapour pressure, $\rho_{l}$ is density of liquid, $u$ is the velocity of the section and $p$ is the outlet pressure. As in Eq \ref{eq:cav_number}, the lower the pressure force in the numerator, the probability of cavitation occurring and its resulting mean cavity length will increase. Thus, to obtain the same mean cavity length as seen in the experiments, the pressure at outlet is adjusted for every calculation. This ensures a consistency with the experiments concerning the cavitation process and aids the study to focus on the turbulence modelling solely. There exist some cases where no unsteady cloud cavitation exists across various cavitation numbers. These cases will be discussed in further detail in Section  \ref{results}. The Re number is 1.8 $\times 10^5$ For local comparison, profiles are plotted at 1.5 mm, 3mm, 5mm, 10mm, 15mm and 20 mm downstream shown in Fig \ref{profiles}. Uniform velocity is implemented at the inlet of the computational domain.  

\begin{table}
\caption{\label{tab:table1} Convergence study of time step and mesh grid}
{\begin{tabular}{lccccc} \toprule
 Timestep & Mesh & Cavity length (mm) & Shedding frequency (Hz) & Computing-time (hrs)  & Average $y^{+}$ \\ \midrule
0.00001 & 100 $\times$ 50 & 27 & 118 & 25 &  2.58 \\
0.000001 & 100 $\times$ 50 & 27 & 115 & 29 & 2.58 \\
0.000001 & 200 $\times$ 100 &27 & 122.5 & 116 & 1.046 \\
0.000001 & 300 $\times$ 100 & 27 & 127.5 & 174 & 0.421 \\ \bottomrule
\end{tabular}}
\end{table}

\begin{table}
\caption{\label{tab:table2} Experimental data for comparison with Table \ref{tab:table1}}
{\begin{tabular}{lccc} \toprule
Case & Cavitation number &Cavity length & Shedding frequency\\
\hline
Experimental & 1.15 & 27 mm & 161\\ \bottomrule
\end{tabular}}
\end{table}

\subsection{Numerical methods}

The numerical simulations are conducted using OpenFOAM \citep{weller1998tensorial}, an open-sourced package consisting of several application-driven solvers and has been utilized for modelling various problems in fluid mechanics. The solver used in the calculations in this study is interPhaseChangeFoam, an unsteady, multiphase and isothermal solver.
The calculation is initially run for 0.2 $T_{ref}$ as a single-phase turbulent flow where the vaporization coefficients in the cavitation model are set to zero followed by a sinusoidal ramp  for approximately 0.07 $T_{ref}$; at the start of the ramp, there is no cavitation while at the end of the ramp, we have fully launched cavitation. The fully cavitating flow is implemented for 0.7 $T_{ref}$. The analysis on a global and local scale is focused on the final portion of the calculation. To provide a distinct analysis, a $T_{ref}$ value of 1.428 s is utilized ensuing simulating 1s of fully cavitating flow. 

As the numerical solution is unsteady, a PIMPLE (a hybrid mixture of the standard PISO and SIMPLE algorithms) algorithm solver is employed for the simulation. A transient, first order implicit method is used for time discretization. Simulations run by first-order were able to converge faster and were more stable than other schemes like second-order Crank-Nicholson. The convective terms are discretized mostly using the Gauss linear second-order upwind scheme with the exception of volume fraction convective flux term using the Van Leer scheme, a total variation diminishing (TVD) scheme. The Laplacian terms are discretized using a second order limited corrector with a blending factor to make sure the non-orthogonal correction is equal to the orthogonal correction to prevent unboundedness. The surface normal gradients to be evaluated at the faces are also discretized in the same manner. For controlling the boundedness of the volume fraction, a semi-implicit multi-dimensional limiter for explicit solution (MULES) is implemented: an implicit corrector step is first implemented corresponding to the discretization schemes and then an explicit correction is applied with the MULES limiter. 

A time-step and grid independence study is also conducted across three different grids and three fixed value timesteps, shown in Table \ref{tab:table1} using the k-$omega$ SST model with the Reboud correction. As observed, despite changing the time step and grid, the average cavity length is equal to the one obtained in the experiments (Table \ref{tab:table2}). The grids are designed with more cells and further refinement in the diverging section of the venturi in order to maintain the focus on the cavitating region. The approximate computing time for the calculations is also displayed, showing increase in computing time approximately proportional to the grid size. It should be noted that this computing time is only for the full cavitation regime of 0.7 $T_{ref}$ on a 32 processor computing system.We also compare the shedding frequency of the calculations-albeit lower than the experiments, it is in relative close range throughout the four studies. Thus, it can be stated that the setup is independent of grid and time step size.

\section{Numerical results}\label{results}
\subsection{Evolution of cloud cavitation}
\begin{figure}
\centering
{\resizebox*{8cm}{!}{\includegraphics{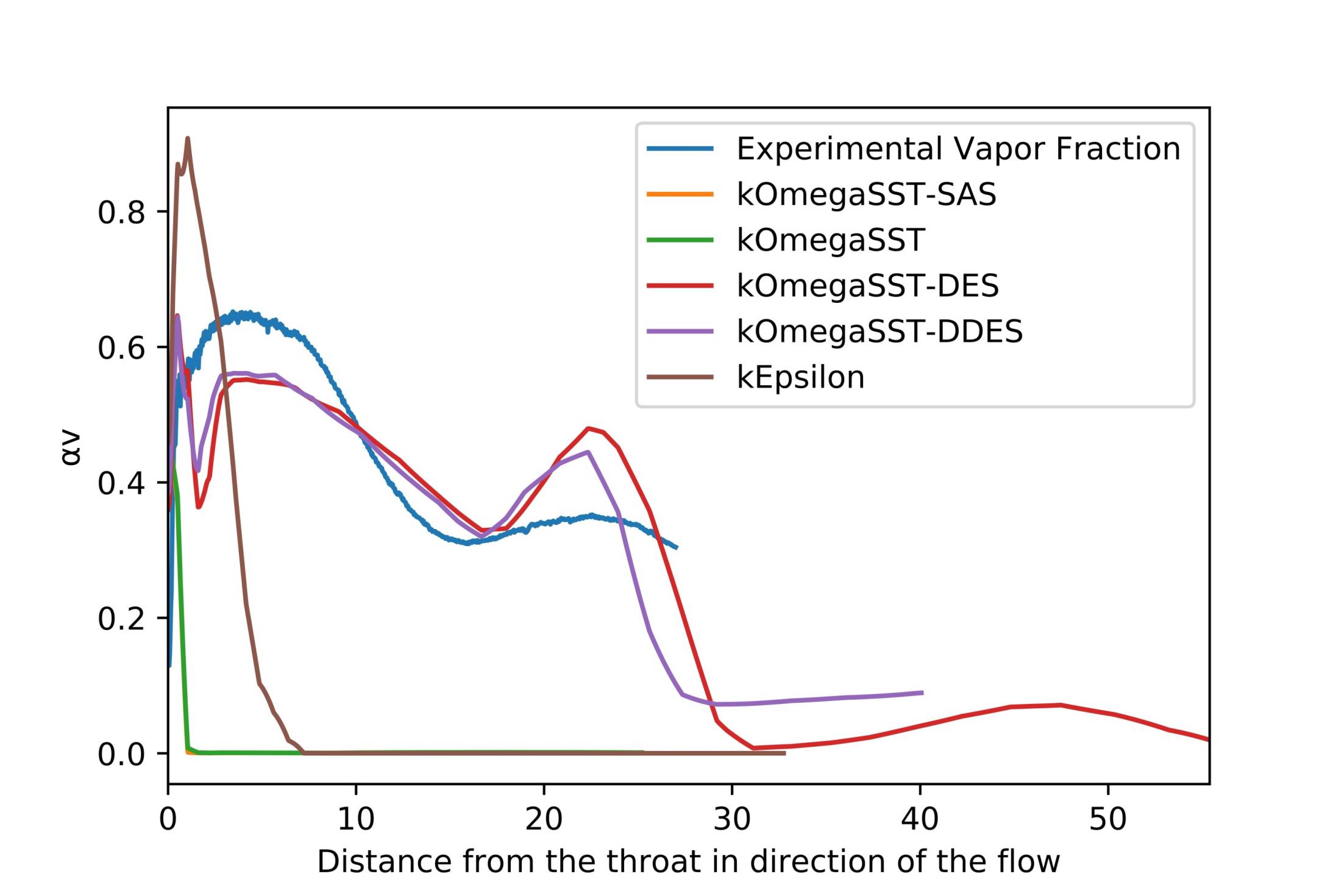}}}\hspace{8pt}
{\resizebox*{8cm}{!}{\includegraphics{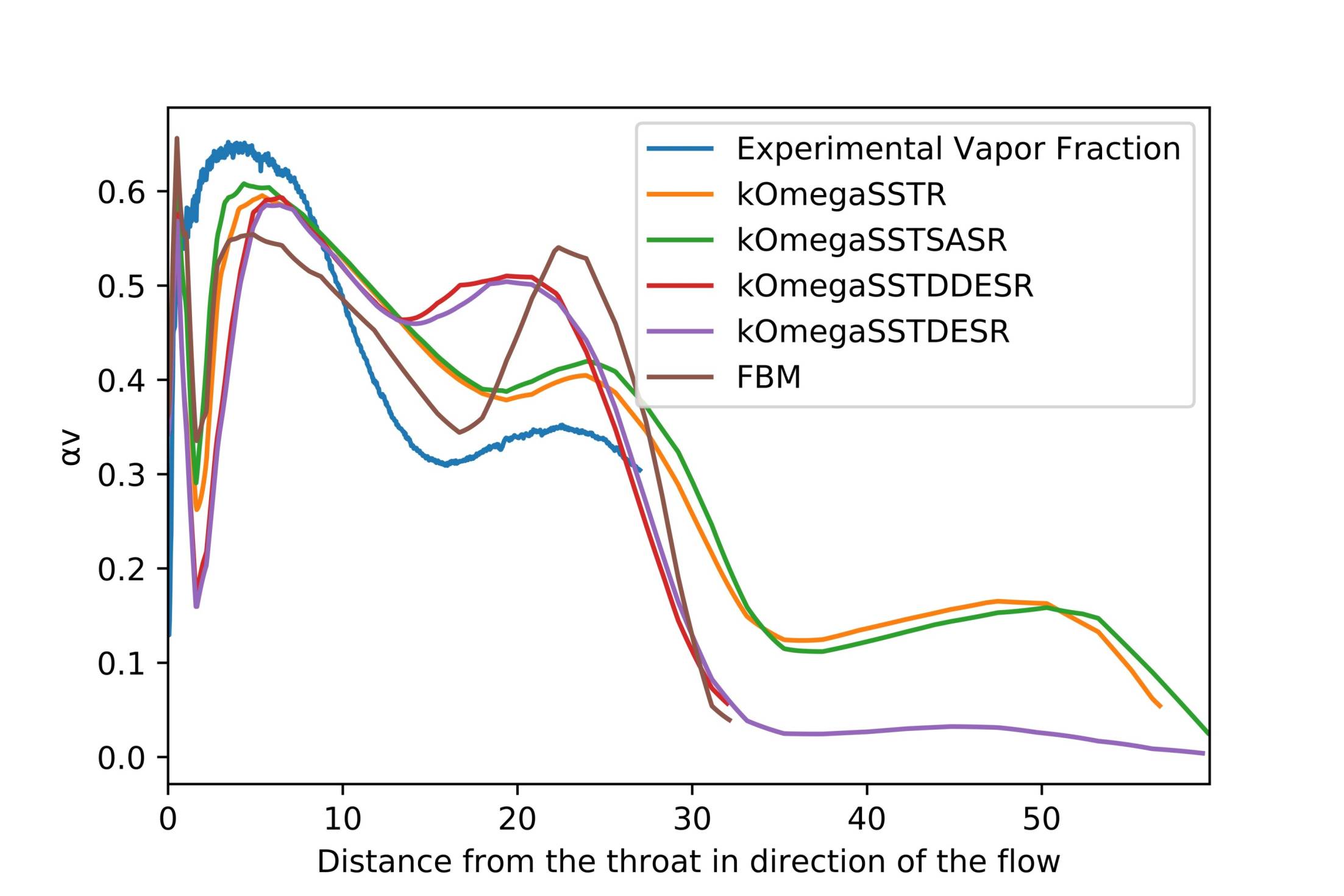}}}
\caption{Mean void fraction plots for a) Standard models and b)Models with Reboud correction} \label{2DC_standard_models}
\end{figure}

In this section, we evaluate and compare the turbulence models with data from the experiment. As stated previously, the outlet pressure is changed to have a cavity length consistent with the experiments. To obtain a general global idea of the unsteadiness, the mean void fraction plot is illustrated. Here, time-averaged void fraction data is plotted over a line splitting the vapor cloud into two equal parts and compared with the grayness level from the images in the experiments 
Fig \ref{2DC_standard_models} (a) shows the mean void fraction plot of the standard models on the coarse mesh. The experimental data shows two bumps- the one close to the origin or the throat is the primary cloud cavity while the one farther downstream is the secondary cloud cavity. The standard k-$\epsilon$, k-$\omega$ SST and the k-$\omega$ SST-SAS models show an initial bump accounted by a cavity formed at the start of the full cavitation regime and vanishing off completely without any secondary detachment. The  k-$\omega$ SST- DES and  k-$\omega$ SST-DDES models show both the primary and secondary cavity albeit a bit more downstream than the experimental data.  

However,once the Reboud correction is implemented, all the models start showing both the primary and secondary cloud detachments (Fig \ref{2DC_standard_models} (b)). An interesting fact to note is that pairs of the corrected  k-$\omega$ SST and the k-$\omega$ SST-SAS models and the k-$\omega$ SST- DES and  k-$\omega$ SST-DDES models  show near-identical behaviours but otherwise it is observed in the latter pair, the secondary cloud detachment occurs a bit more upstream that the former pair and the FBM model for whom, there is not a considerable difference in cloud shedding.  
The cavity evolution plots for some of the models are illustrated. These plots are plotted on a distance from the throat vs time axis color coded by the minimal density in each cross section of the venturi once the full cavitation regime has been launched. Studying the plot on a global basis presents the concepts of periodic vapor shedding and shedding frequency. 

\begin{figure}
\centering
{\resizebox*{8cm}{!}{\includegraphics{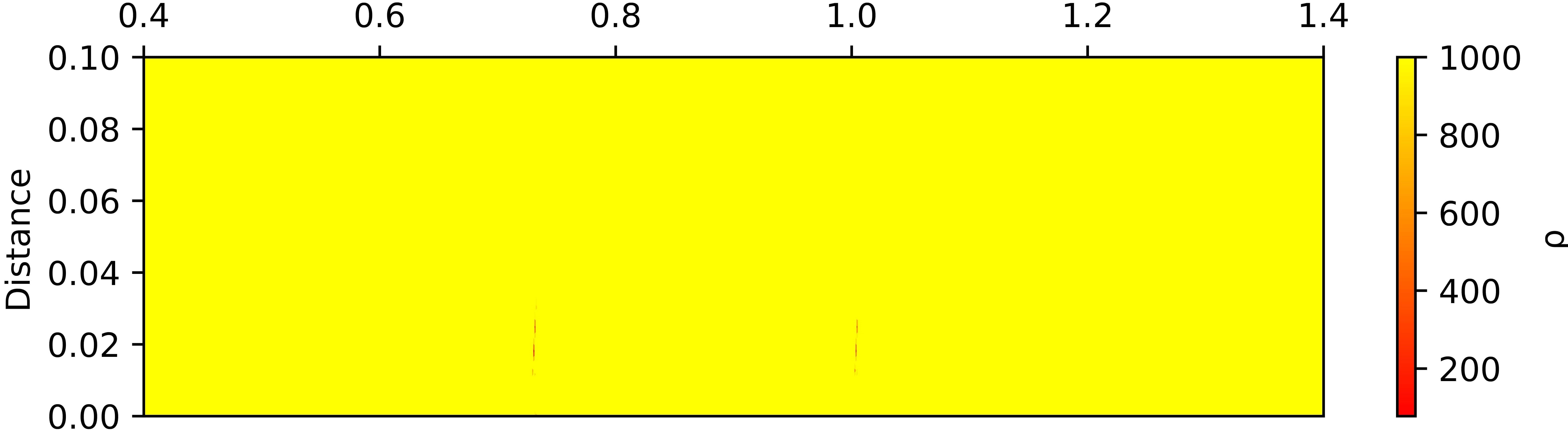}}}\hspace{8pt}
{\resizebox*{8cm}{!}{\includegraphics{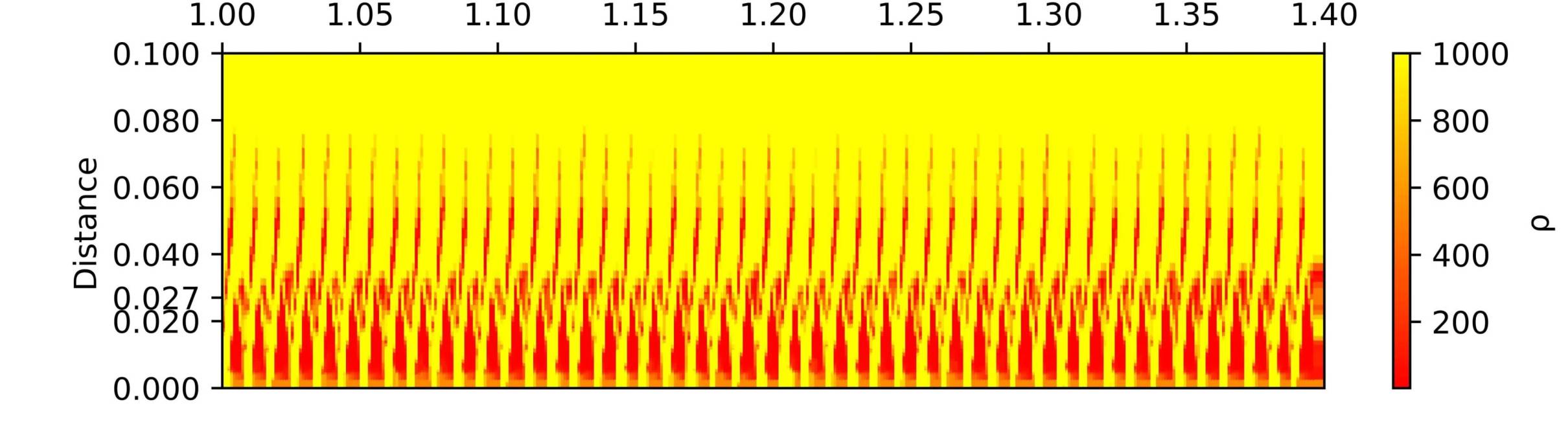}}}
\caption{Cavity evolution plots for a)Standard k-$\omega $-SST and b) k-$\omega $-SST with Reboud correction for coarse mesh. Distance in mm}
\label{2D_coarse_kOmegaSST}
\end{figure}

Looking at Fig \ref{2D_coarse_kOmegaSST} (a) for the k-$\omega$-SST simulation for the coarse mesh , no cloud cavity shedding is observed at all. Comparing it with the calculation of the same model but implemented with Reboud correction (Fig \ref{2D_coarse_kOmegaSST} (b)), we see periodic vapor shedding throughout the simulation.  Each of these shedding cycles consists of two phases: the first phase from the bottom of the X-axis to approximately 27 mm is the primary cloud cavity that detaches from the venturi throat followed by the second phase slightly askew stretching till 80 mm. This phase is the secondary cloud detachment downstream. 

\begin{figure}
\centering
{\resizebox*{8cm}{!}{\includegraphics{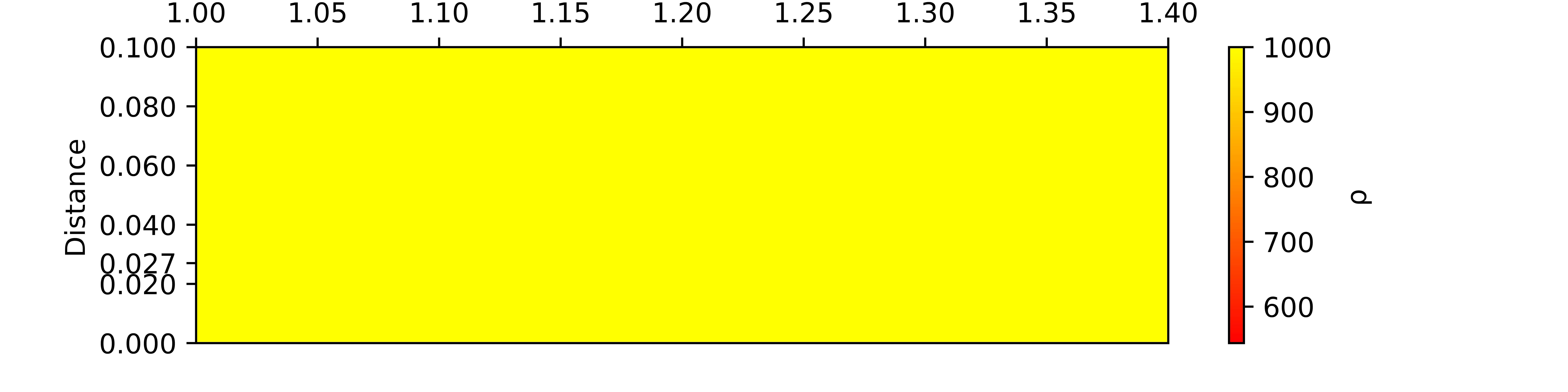}}}\hspace{8pt}
{\resizebox*{8cm}{!}{\includegraphics{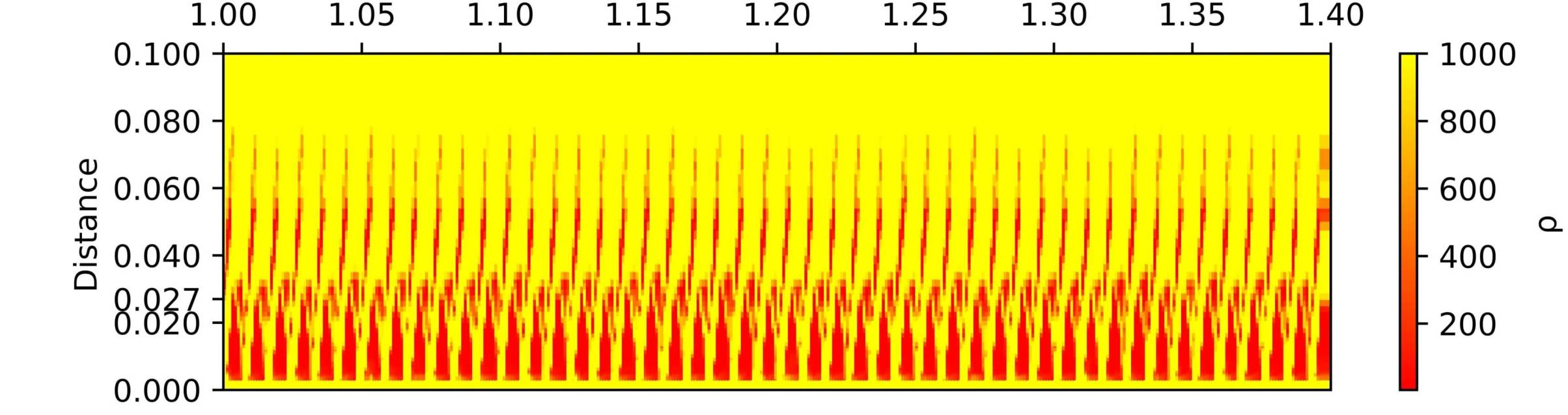}}}
\caption{Cavity evolution plots for a)Standard k-$\omega $ SST-SAS and b) k-$\omega$ SST-SAS with Reboud correction for coarse mesh. Distance in mmn} \label{2D_coarse_kOmegaSSTSAS}
\end{figure}

\begin{figure}
\centering
{\resizebox*{8cm}{!}{\includegraphics{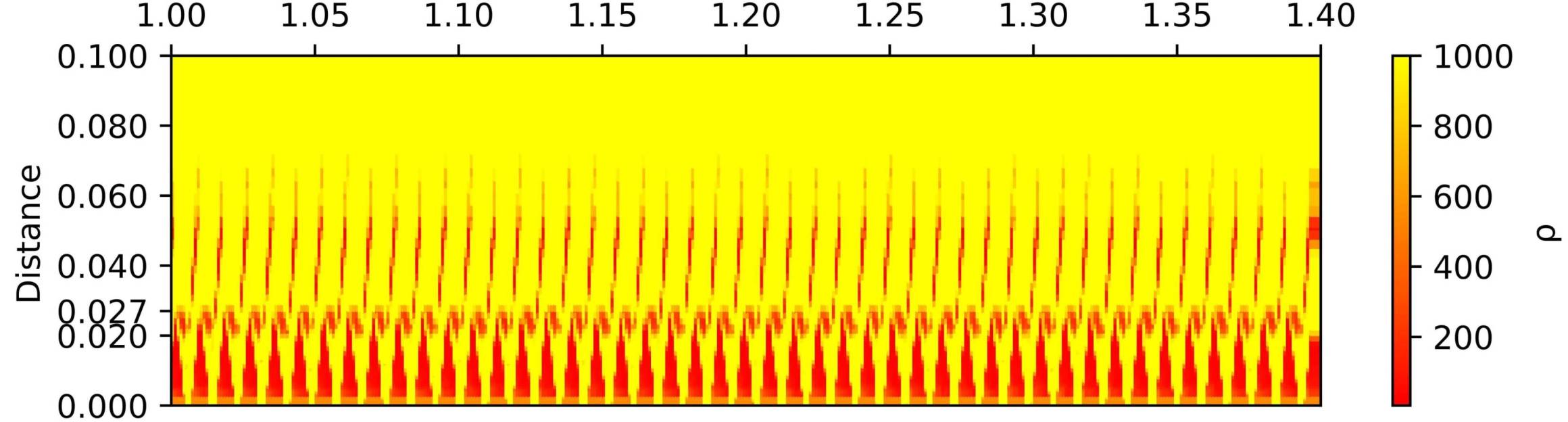}}}\hspace{8pt}
{\resizebox*{8cm}{!}{\includegraphics{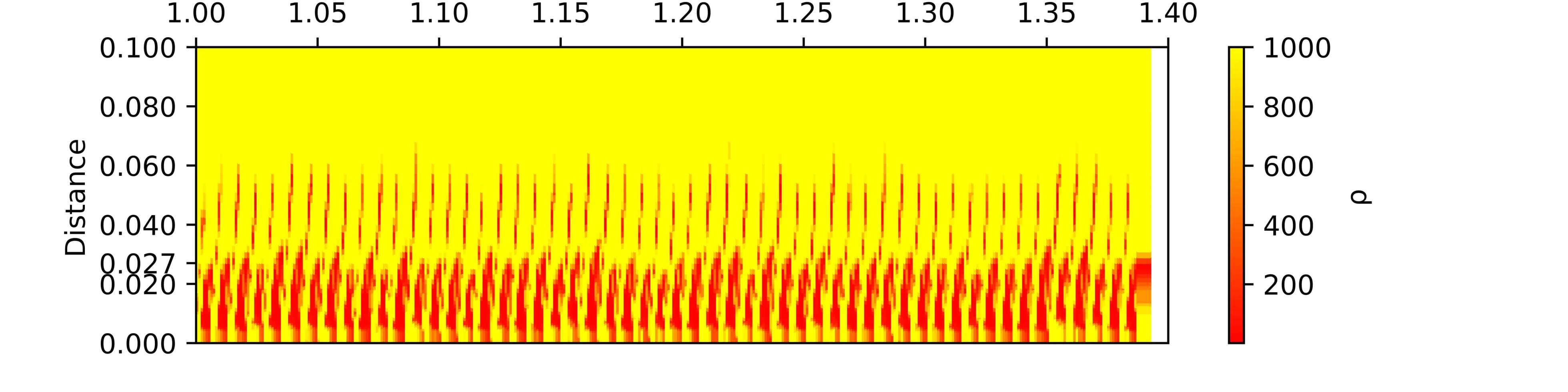}}}
\caption{Cavity evolution plots for a)k-$\omega $ SST-DES and b) k-$\omega$ SST-DES with Reboud correction for coarse mesh. Distance in mm} 
\label{2D_coarse_kOmegaSSTDES}
\end{figure}

A similar observation is noted when the cavity evolution plots for the k-$\omega$ SST-SAS model with and without the Reboud correction are plotted. However, on plotting the cavity evolution for the k-$\omega$ SST-DES model with and without the Reboud correction, both the models show identical behaviour. 
\begin{figure}
\centering
{\resizebox*{8cm}{!}{\includegraphics{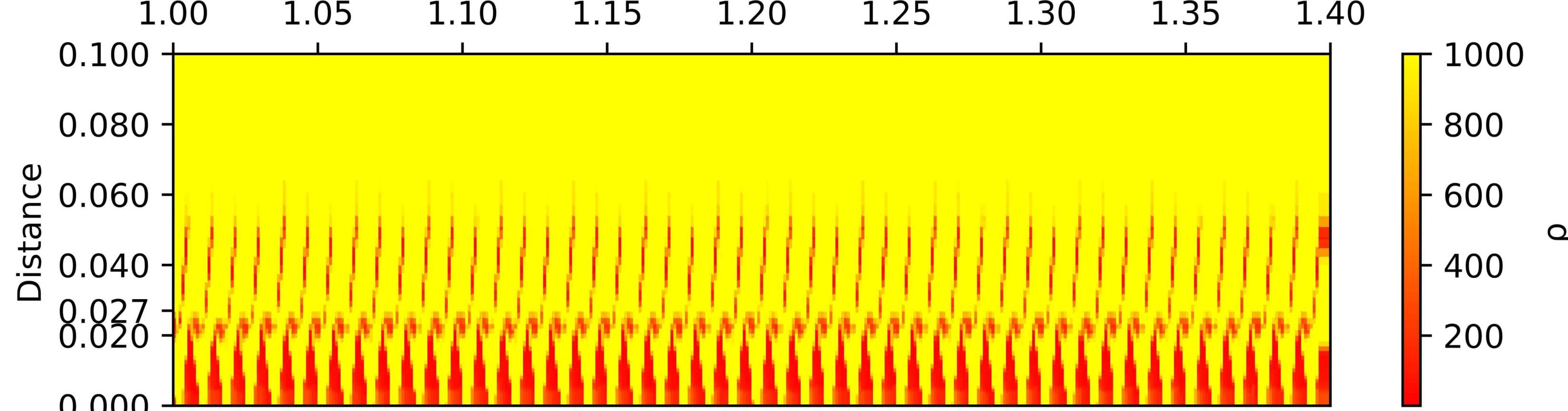}}}\hspace{8pt}
{\resizebox*{8cm}{!}{\includegraphics{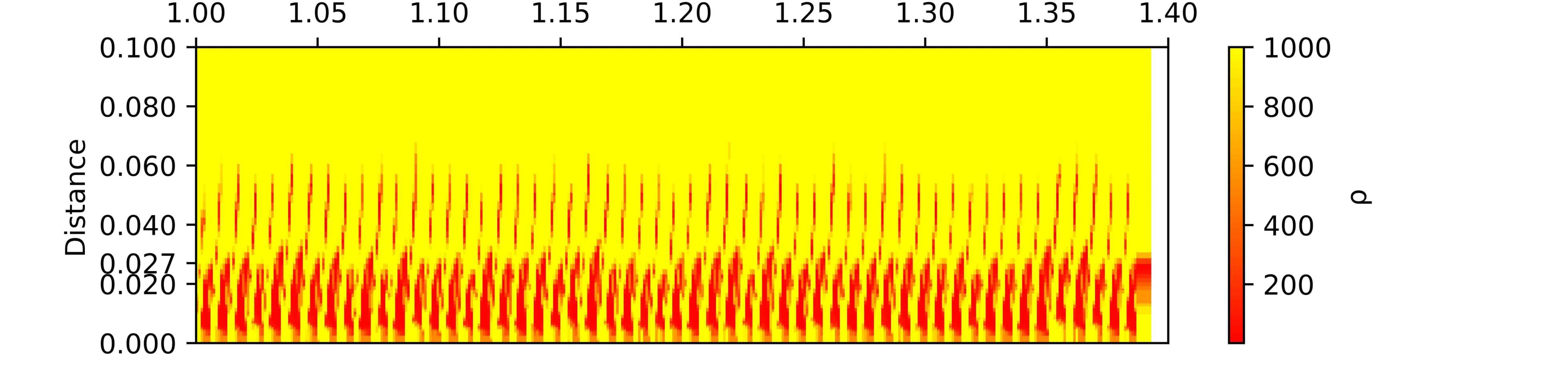}}}
\caption{Cavity evolution plots for a)Standard k-$\omega $ SST-DDES and b) k-$\omega$ SST-DDES with Reboud correction for coarse mesh. Distance in mm} \label{2D_coarse_kOmegaSSTDDES}
\end{figure}

\begin{figure}[htb]
\centering
\includegraphics[width=10cm, height=2cm]{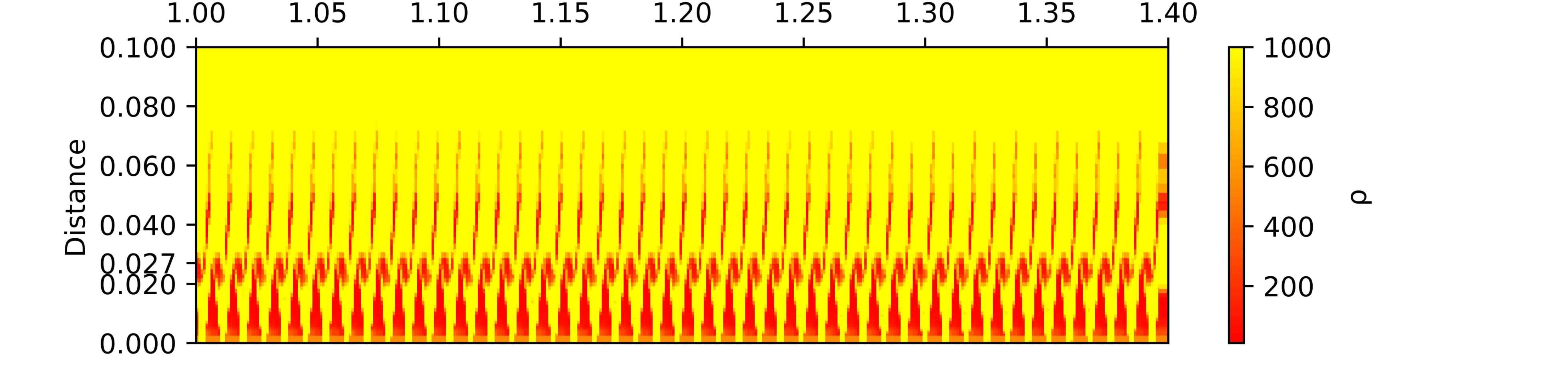}
\caption{Cavity Evolution Plot for the FBM model for the coarse mesh. }
\label{2D_coarse_FBM}
\end{figure}

Using the Delayed DES (DDES), an improved version of the DES model also yields identical results along with the Reboud correction implemented (see Fig \ref{2D_coarse_kOmegaSSTDDES} (a) and \ref{2D_coarse_kOmegaSSTDDES}(b)). A similar observation is noted when the cavity evolution plot for the FBM model is plotted (Fig \ref{2D_coarse_FBM}). In all of the above-mentioned cases, we see turbulence modelss with the Reboud correction are able to predict unsteady cloud cavity shedding with a regular primary cavity length of 27 mm as seen in the experiments followed by a secondary cloud detachment . It can also be observed that some hybrid RANS-LES models like the DES and DDES models are able to show the unsteady vapor shedding without the need of the Reboud correction.  

\begin{figure}
\centering
{\resizebox*{8cm}{!}{\includegraphics{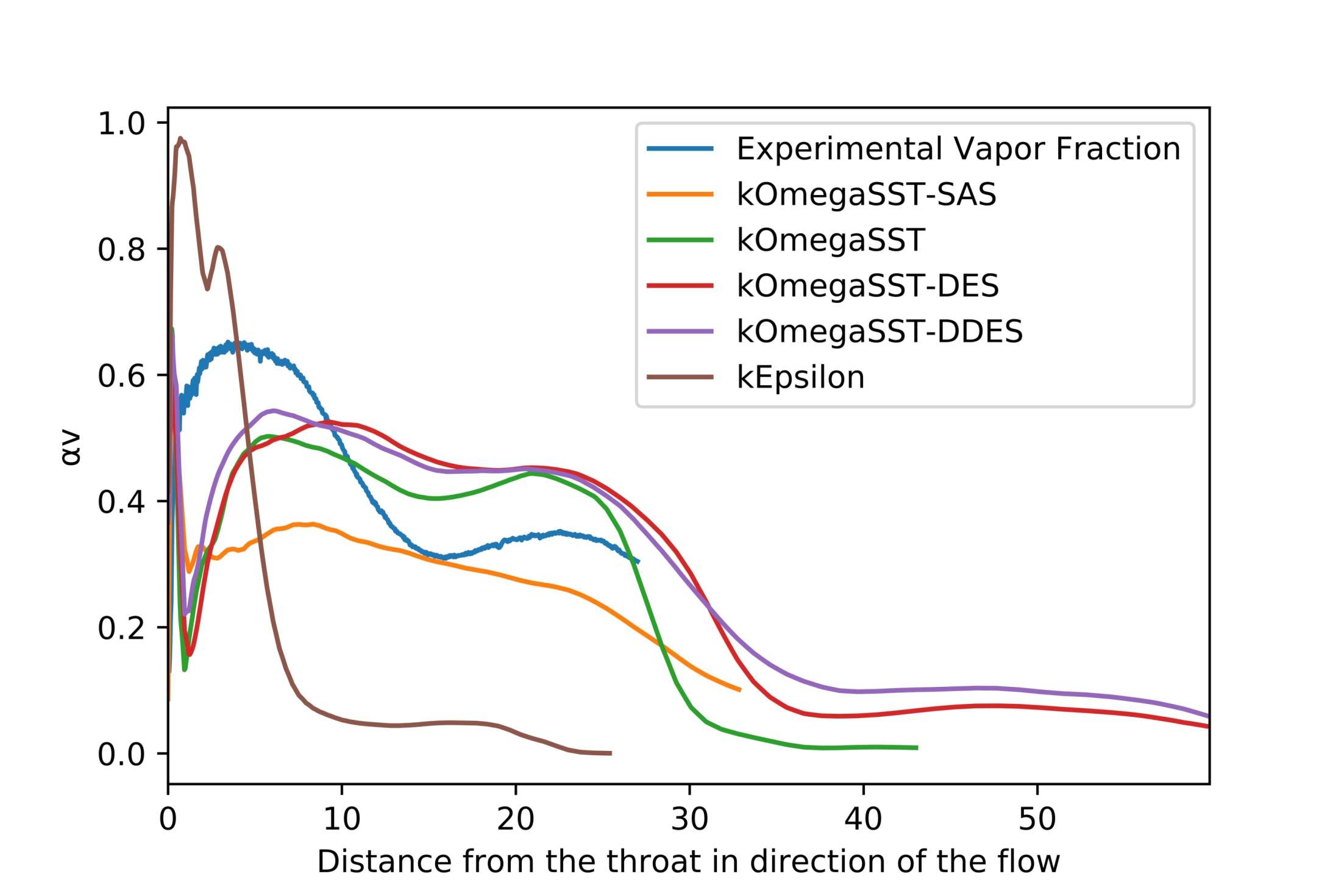}}}\hspace{8pt}
{\resizebox*{8cm}{!}{\includegraphics{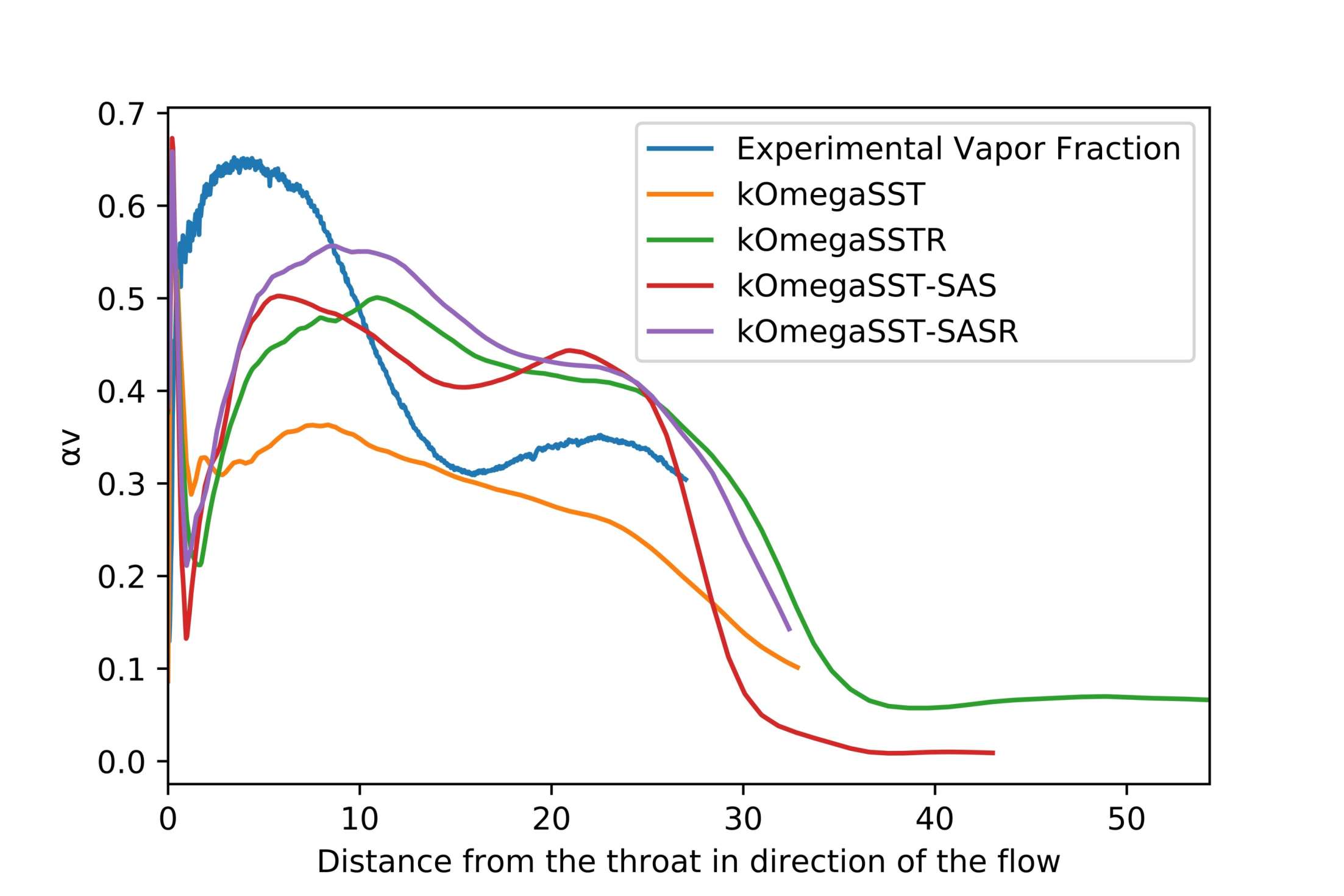}}}
\caption{Mean void fraction plot on intermediate mesh for a)Standard models and b) Models with Reboud correction.} \label{2DI_standard_models}
\end{figure}

A significantly unique development is observed when plotting the time-averaged void fraction plots and the cavity evolution plots for the standard turbulence model simulations on the intermediate and the fine mesh. Fig \ref{2DI_standard_models} (a) shows the time-averaged void fraction plot for the standard models: almost all the standard models are able to show unsteady shedding with the primary and secondary detachment caused due to the collision between the re-entrant jet and the main side-entrant jets. The k-$\epsilon$ model is still unable to predict correctly the cavitation while the standard k-$\omega$ SST model simulation is not able to predict pinch-off of the primary shedding and the resulting secondary cloud detachment. The models are then compared with the models with the Reboud correction models (Fig \ref{2DI_standard_models} (b)). For the sake of clarity. the plot shows only the  k-$\omega$ SST and k-$\omega$ SST-SAS with their corrected versions. Although there is a considerable difference in the pinch-off distance and the distance from the throat where the secondary detachment formation occurs, all the four models do predict unsteady vapor shedding. The cavity evolution plots would further confirm these predictions.   

\begin{figure}
\centering
{\resizebox*{8cm}{!}{\includegraphics{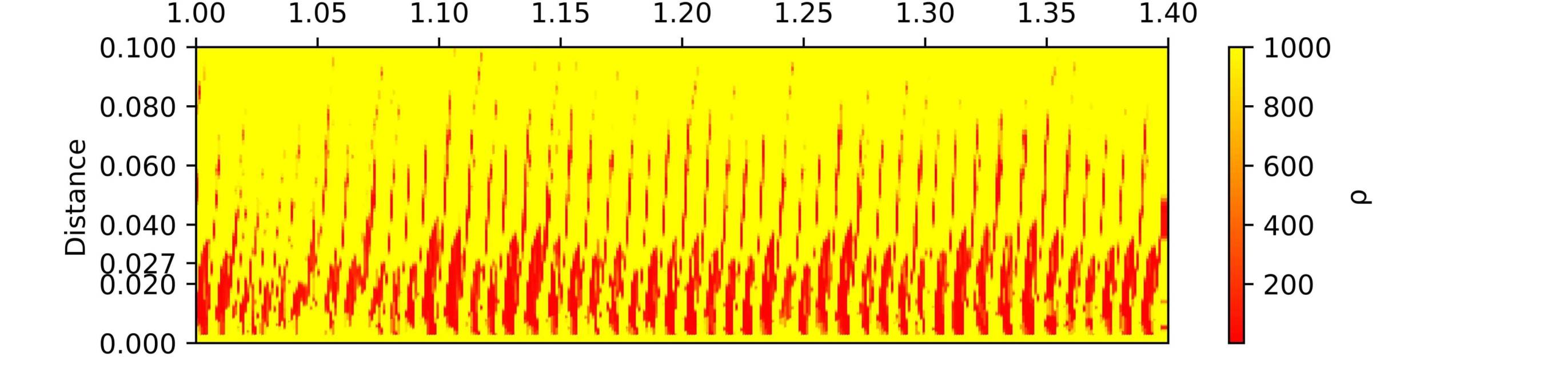}}}\hspace{8pt}
{\resizebox*{8cm}{!}{\includegraphics{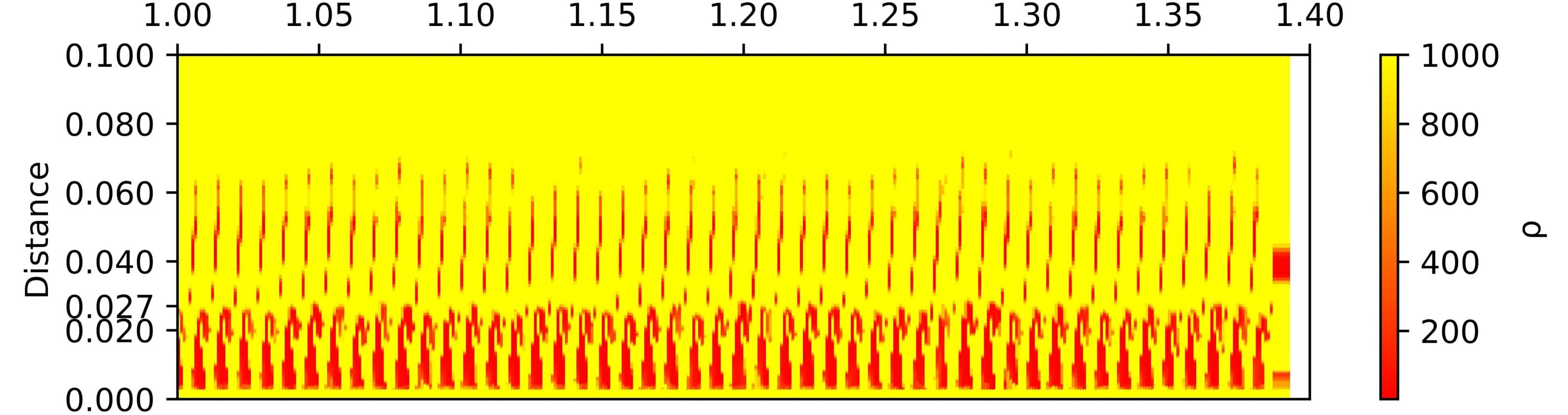}}}
\caption{Cavity evolution plots for a)Standard k-$\omega $ SST and b) k-$\omega$ SST-SAS for intermediate mesh.} \label{2DI_kOmegaSST}
\end{figure}

\begin{figure}[htb]
\centering
\includegraphics[width=10cm, height=2cm]{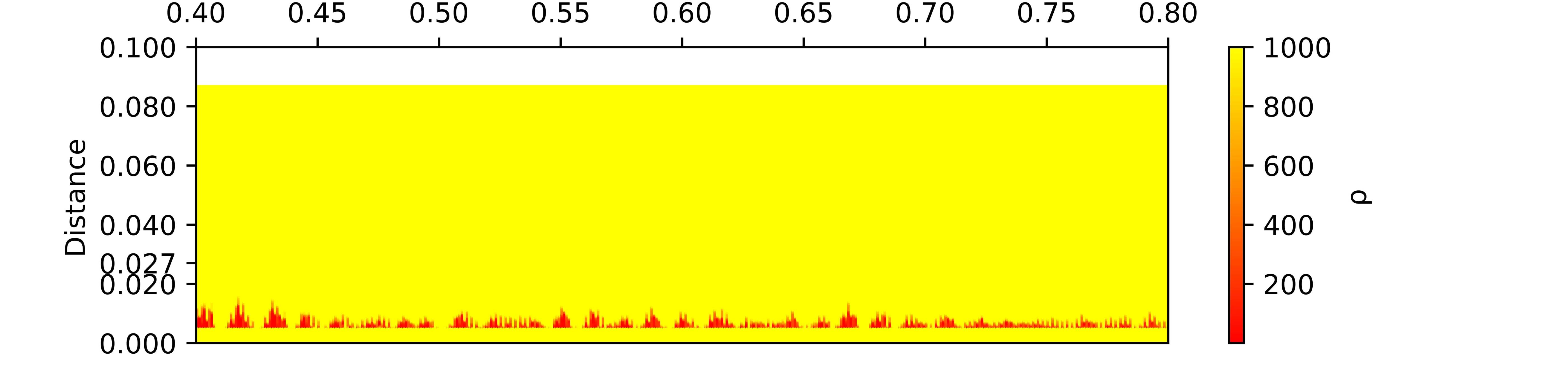}
\caption{Cavity Evolution Plot for the k-$\epsilon$ simulation for the intermediate mesh.}
\label{2DI_kEpsilon} 
\end{figure}

From the cavity evolution plots, the standard k-$\omega$-SST model shown in Fig \ref{2DI_kOmegaSST} (a) is able to predict unsteady cloud cavity similar to the corrected model. While the simulation does not predict periodic unsteady shedding like the corrected one, it still shows some form of unsteady shedding and is able to produce an average cavity length equivalent to the experiments. This is a novel breakthrough as previous studies were unable to reproduce unsteady cloud cavitation using standard RANS models. It could be posited that the possible reasons include the high dissipation rate of the solver leading to a lower over-prediction of the global eddy viscosity field and computational setup where a no cavitation regime and a ramp has been implemented first instead of directly launching a fully cavitating regime immediately. Further, the cavity evolution plot for the k-$\omega$ SST-SAS model(Fig \ref{2DI_kOmegaSST} (b)) shows identical behaviour. However, the k-$\epsilon$ model (see Fig \ref{2DI_kEpsilon}) is still unable to predict any cloud cavitation. Small cavities are visible throughout the regime however, there is no main cavity and thus no secondary either. The other models do show unsteady cloud vapor shedding. Our further local analysis will thus be on the intermediate and fine mesh calculations

\begin{figure}
\centering
{\resizebox*{8cm}{!}{\includegraphics{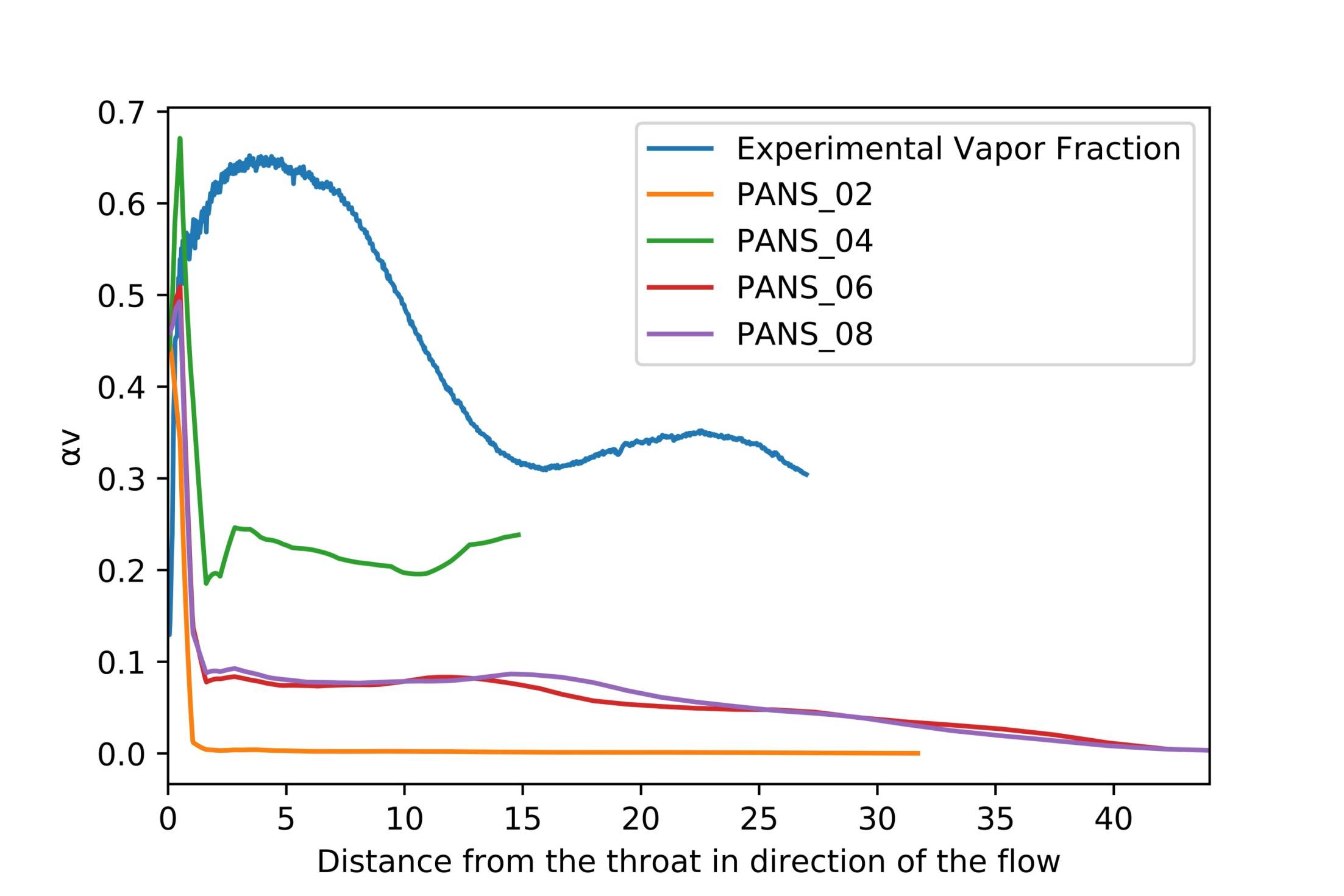}}}\hspace{8pt}
{\resizebox*{8cm}{!}{\includegraphics{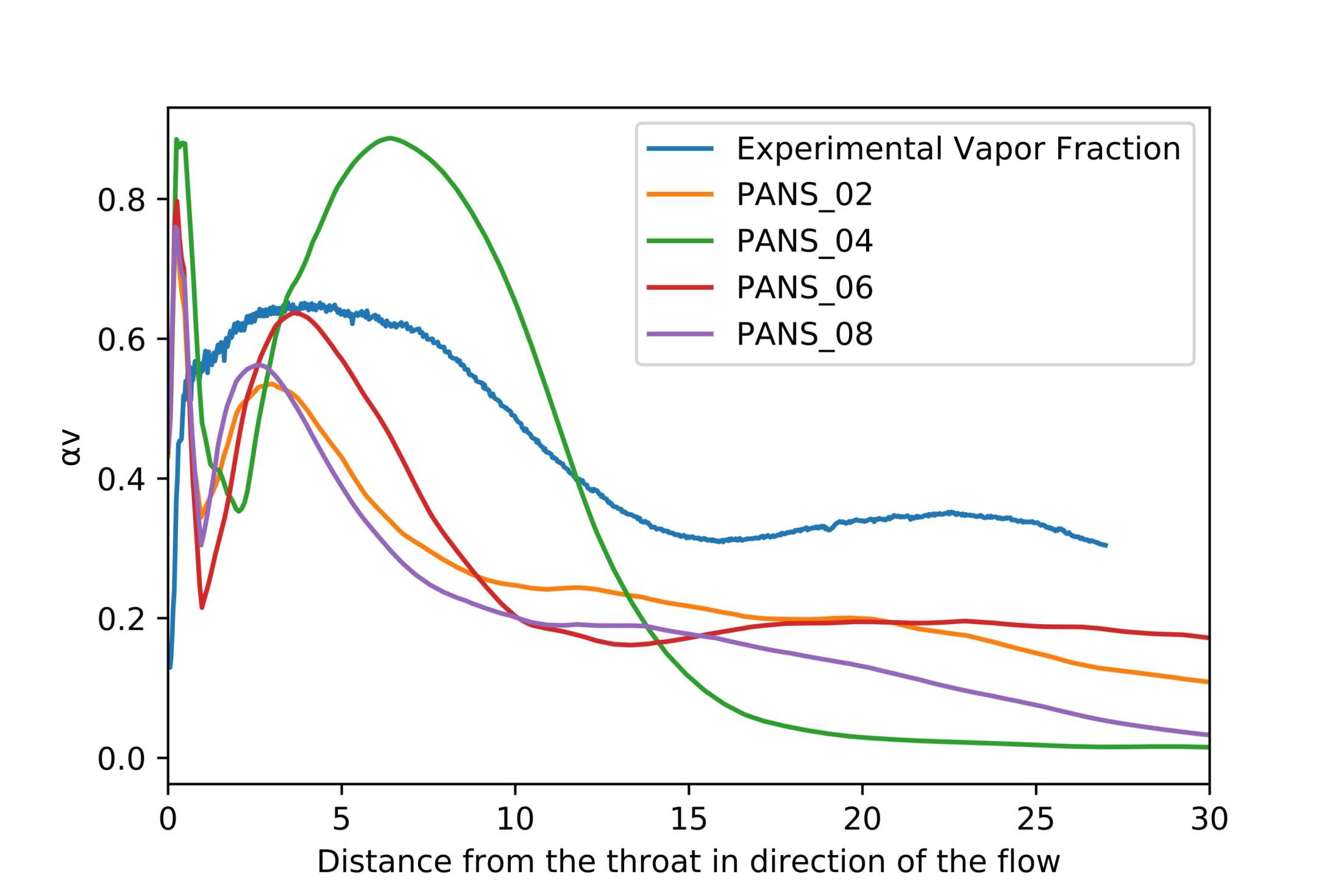}}}
\caption{Mean void fraction plot for PANS simulations with fixed $f_{k}$ on a) coarse mesh and b) intermediate mesh} \label{2DC_PANS}
\end{figure}

The mean void fraction plots for PANS models with fixed $f_{k}$ models for the coarse and intermediate mesh calculations are shown in Figures \ref{2DC_PANS} (a) and \ref{2DC_PANS} (b). The coarse mesh simulations show results identical to the standard model calculations where there is no unsteady vapor shedding captured at all. This would seem intuitive for $f_{k}$ = 0.6 and 0.8 respectively considering the RANS model would dominate the flow characteristics in those cases and concluding earlier from mean void fraction plots for the k-$\omega$ SST model that showed similar results. The PANS model case with $f_{k}$=0.4 does show a slight bump , indicating the formation of a primary cavity but it is extremely small and lacks the secondary detached cavity. The intermediate mesh does show the formation of the main cloud cavity from the sheet very close to the throat distance in the experiments but there is no formation of secondary sheet cavity. This could stem from the fact that PANS model calculations are not able to capture the re-entrant jet formation and thus the resultant secondary cavity. 
It may be considered that the grid may not be fine enough for an accurate PANS calculation. However, Girimaji \textit{et al.}. \citep{girimaji2005partially, girimaji2006partially} have repeatedly stated that the model is grid-independent. Further analysis may be needed where the filter ratio could be varied spatially as a factor of the turbulent length scale on a finer mesh to resolve more structures instead of fixing it throughout the calculation.

\begin{figure}[htb]
\centering
\includegraphics[width=8cm, height=5cm]{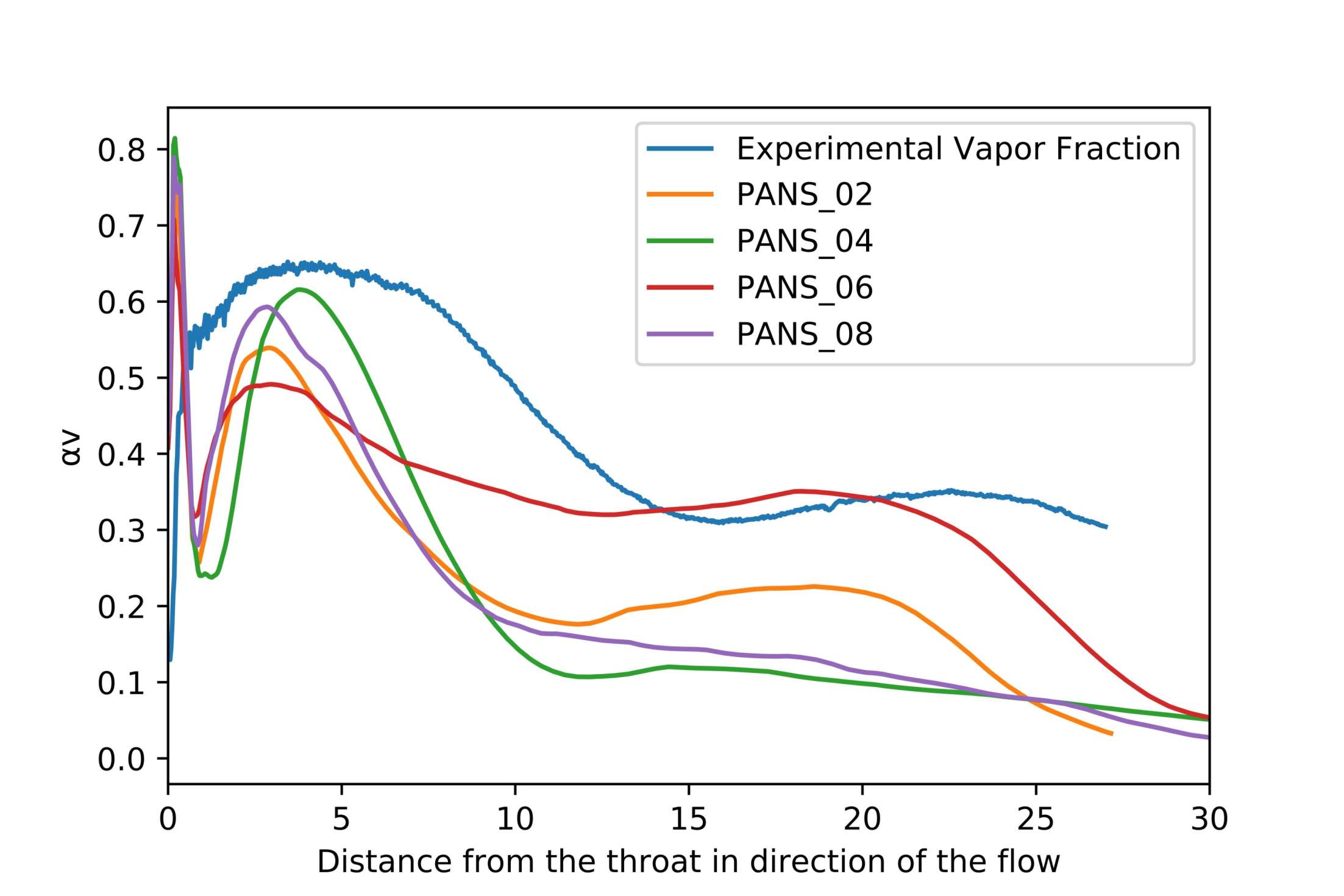}
\caption{Mean void fraction plot of PANS simulations with fixed $f_{k}$ for the fine mesh }
\label{2DF_PANS} 
\end{figure}

\begin{figure}
\centering
{\resizebox*{8cm}{!}{\includegraphics{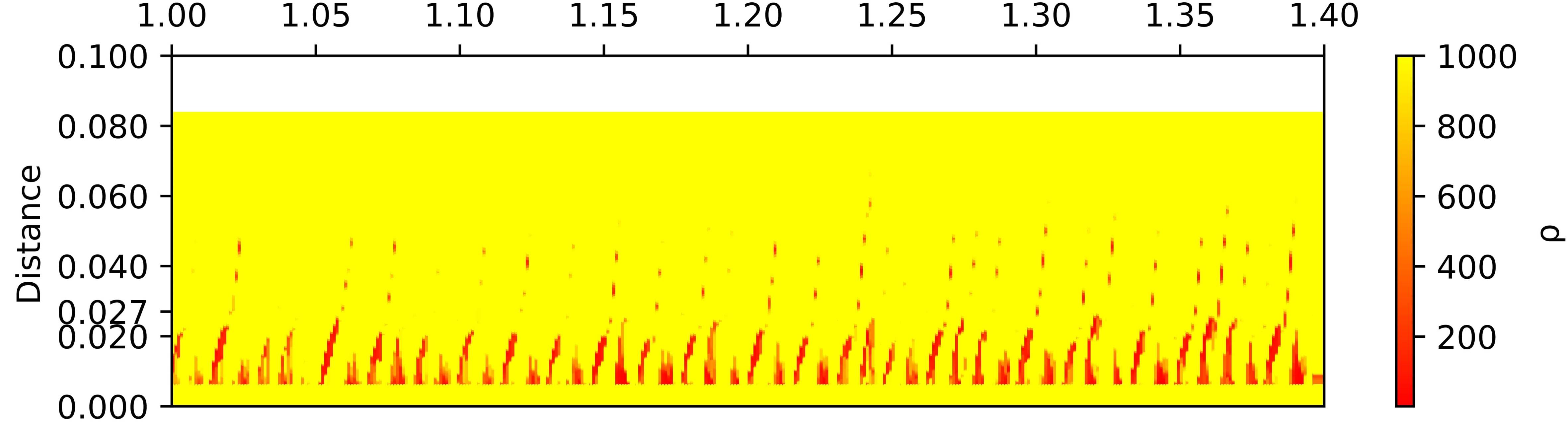}}}\hspace{8pt}
{\resizebox*{8cm}{!}{\includegraphics{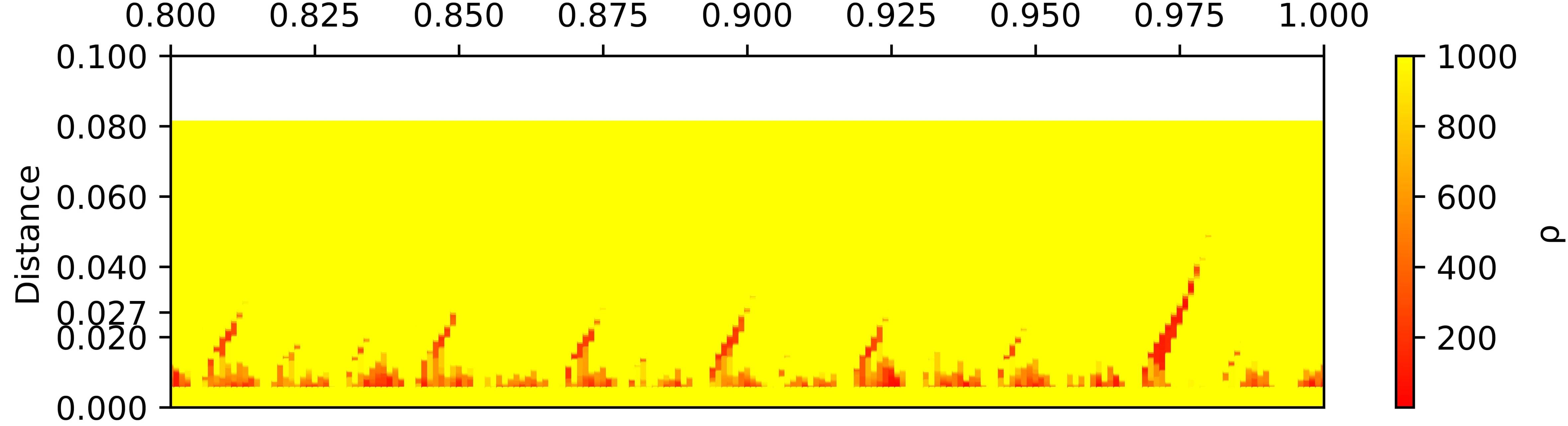}}}
\caption{Cavity evolution plots for PANS simulation with a) $f_{k}=0.2$ for the fine mesh and b) $f_{k}=0.4$ for the fine mesh} \label{2DF_PANS_02}
\end{figure}

\begin{figure}
\centering
{\resizebox*{8cm}{!}{\includegraphics{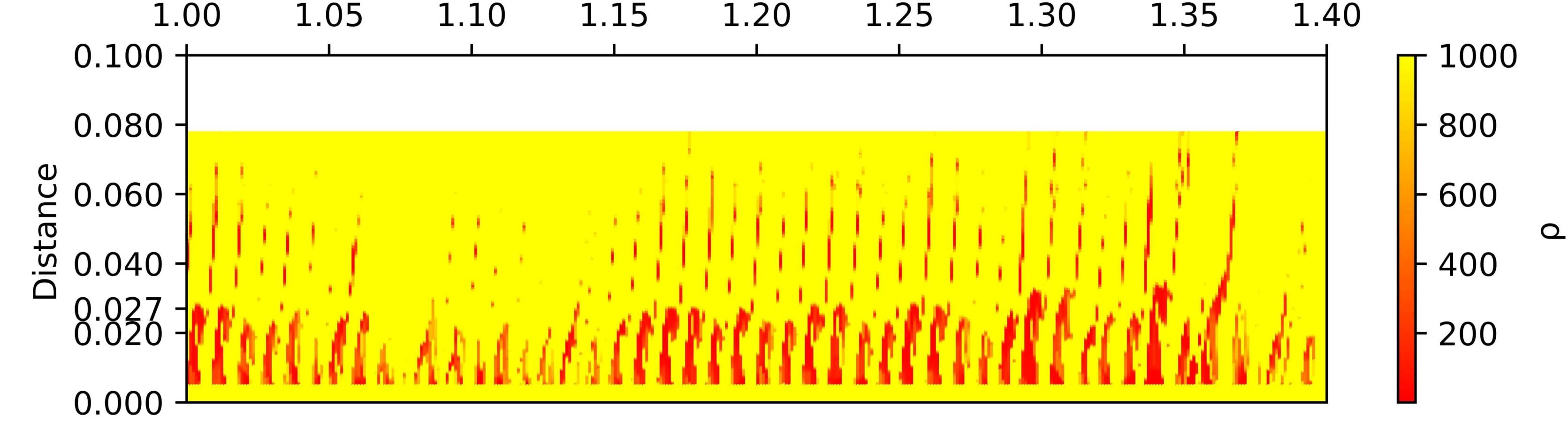}}}\hspace{8pt}
{\resizebox*{8cm}{!}{\includegraphics{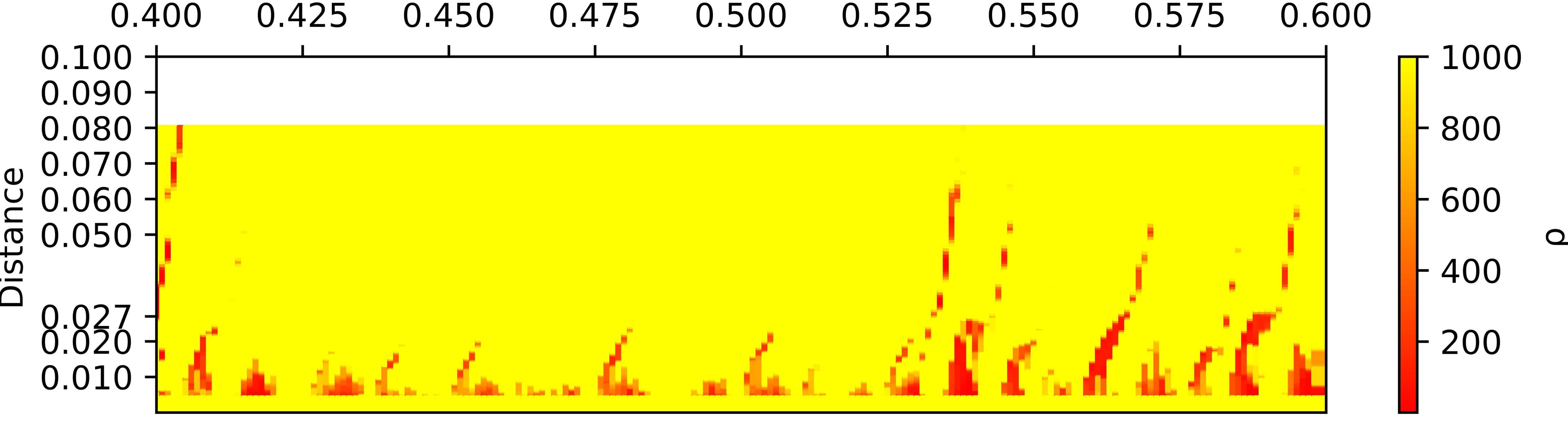}}}
\caption{Cavity evolution plots for PANS simulation with a) $f_{k}=0.6$ for the fine mesh and b) $f_{k}=0.8$ for the fine mesh} \label{2DF_PANS_06}
\end{figure}

We then compare the mean void fraction plot for the PANS simulations on the fine mesh to investigate if the shedding dynamics are captured on the fine mesh. In the Fig \ref{2DF_PANS} , it is observed that the calculation  with $f_{k}=0.2$ shows two bumps as expected- the unresolved kinetic energy component has an extremely low value resulting in more resolved kinetic energy components. The simulation with $f_{k}=0.6$ shows a similar cloud cavity development but the simulations with $f_{k}=0.4$ and $f_{k}=0.8$  are able to predict only a primary cloud cavity and no resulting secondary cavity. It can be considered that either the secondary cavity is either too small to be captured or the simulations are not able to capture the re-entrant jet and thus the secondary cavity as in the simulations on the intermediate mesh. The cavity evolution plots for the four simulations are able to demonstrate that clearly. Looking at Figures \ref{2DF_PANS_02} (a) and \ref{2DF_PANS_06} (a), with $f_{k}=0.2$ and $0.6$, periodic vapor shedding is clearly seen unlike \ref{2DF_PANS_06} (b) and \ref{2DF_PANS_02} (b) where some vapor shedding cycles are observed but they are not periodic. Table \ref{tab:table4} shows the shedding frequencies of the simulations specified above. As compared to Table \ref{tab:table3}, although the models are able to obtain the same mean cavity length in several cases, none of them have the same shedding frequency as the experiments. This amplifies the influence of the turbulence model on the cavitation modelling on the global scale. 

\begin{table}
\caption{\label{tab:table4} Shedding frequencies (Hz) of simulations on the three meshes. Here, 'R' denotes the Reboud correction}
{\begin{tabular}{lccc} \toprule
 Model & Coarse mesh & Intermediate mesh & Fine mesh \\ \midrule
kEpsilon & 0 & 0 & 0 \\
kOmegaSST & 0 & 125 & 125 \\
kOmegaSSTR & 115 & 135 & 135 \\
kOmegaSSTSAS & 0 & 120 & 135 \\
kOmegaSSTSASR & 117.5 & 130 & 135 \\ 
kOmegaSSTDES & 112.5 & 125 & 125 \\
kOmegaSSTDDES & 115 & 120 & 125\\
FBM & 115 & 115 & 120 \\
PANS ($f_{k}=0.2$) & 0 & 15 & 57.5 \\ 
PANS ($f_{k}=0.4$) & 0 & 0 & 0 \\ 
PANS ($f_{k}=0.6$) & 0 & 0 &  85\\ 
PANS ($f_{k}=0.8$) & 0 & 0 & 0 \\ \bottomrule
\end{tabular}}
\end{table}

\subsection{Predicting turbulence on a local scale}

\begin{figure}
\centering
{\resizebox*{15cm}{!}{\includegraphics{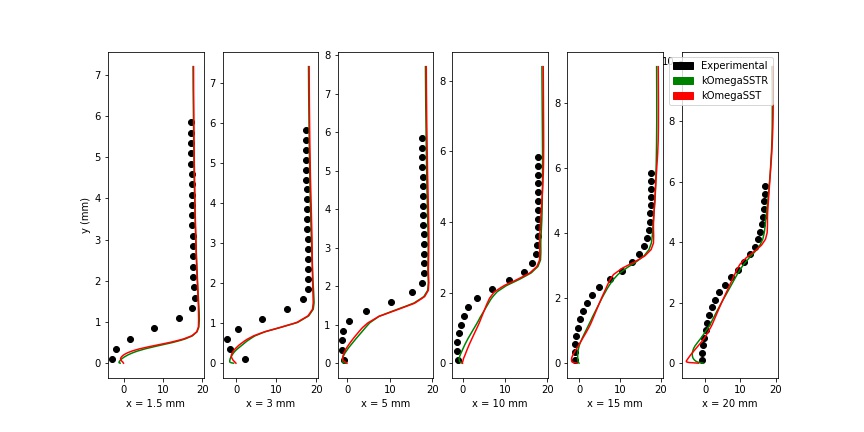}}}\hspace{8pt}
{\resizebox*{15cm}{!}{\includegraphics{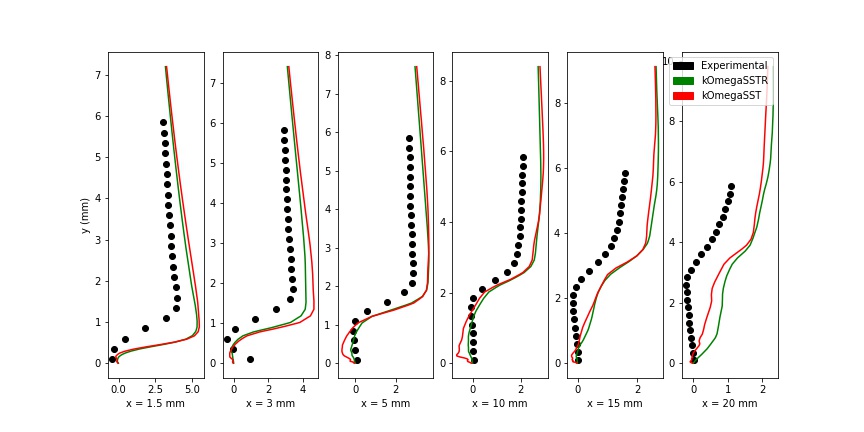}}}
\caption{Comparison for time-averaged velocity in both a) stream wise direction and b) wall direction} \label{UMeanX_SST}
\end{figure}

\begin{figure}
\centering
{\resizebox*{15cm}{!}{\includegraphics{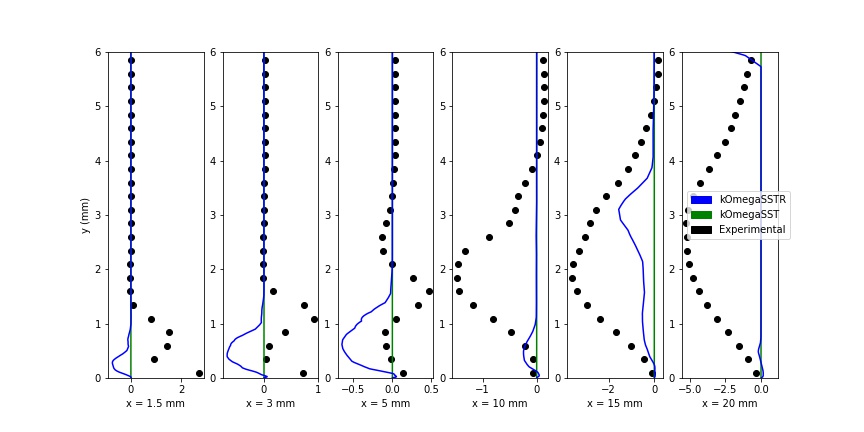}}}\hspace{8pt}
{\resizebox*{15cm}{!}{\includegraphics{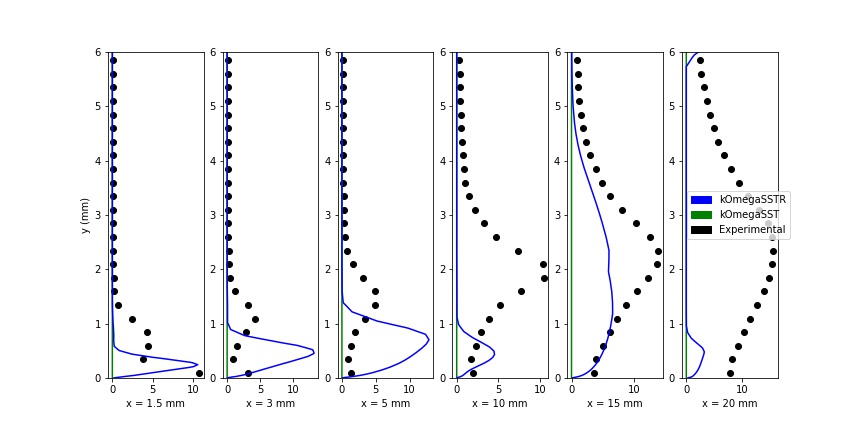}}}
\caption{Local comparisons between the experiments, standard k-$\omega$ SST calculation and k-$\omega$ SST calculation with the Reboud correction on the intermediate mesh for a) Reynolds shear stress and b) Turbulent Kinetic energy} \label{Tau12_SST}
\end{figure}

\begin{figure}
\centering
{\resizebox*{15cm}{!}{\includegraphics{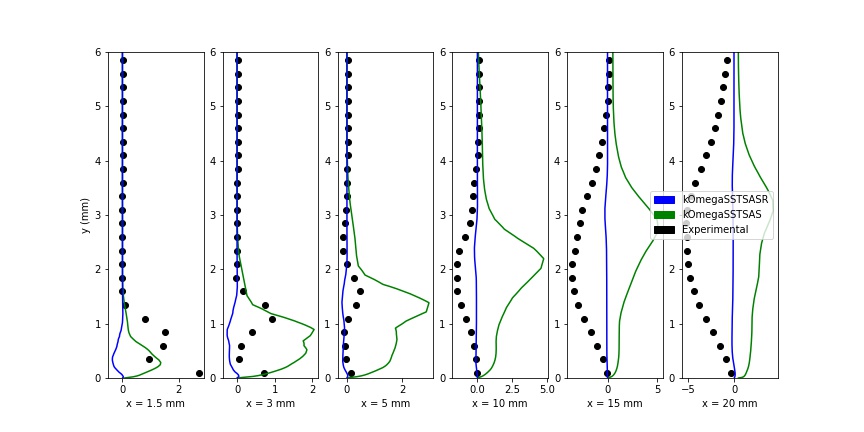}}}\hspace{8pt}
{\resizebox*{15cm}{!}{\includegraphics{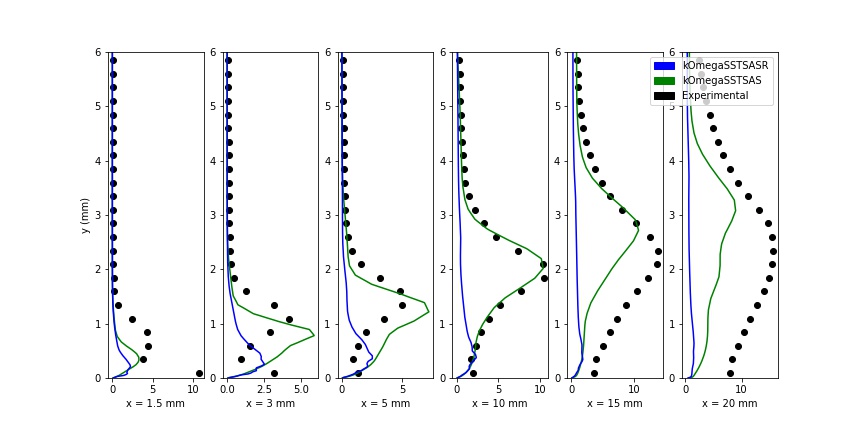}}}
\caption{Local comparisons between the experiments, standard k-$\omega$ SSTSAS calculation and k-$\omega$ SSTSAS calculation with the Reboud correction on the intermediate mesh for a) Reynolds shear stress and b) Turbulent Kinetic energy} \label{Tau12_SAS}
\end{figure}

\begin{figure}
\centering
{\resizebox*{15cm}{!}{\includegraphics{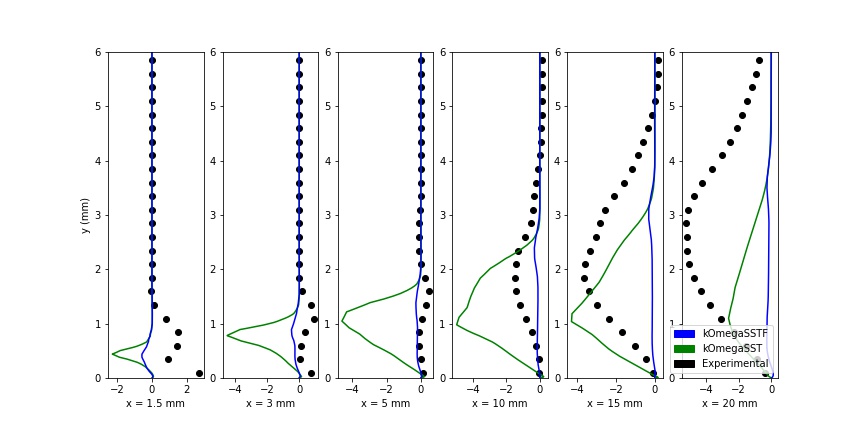}}}\hspace{8pt}
{\resizebox*{15cm}{!}{\includegraphics{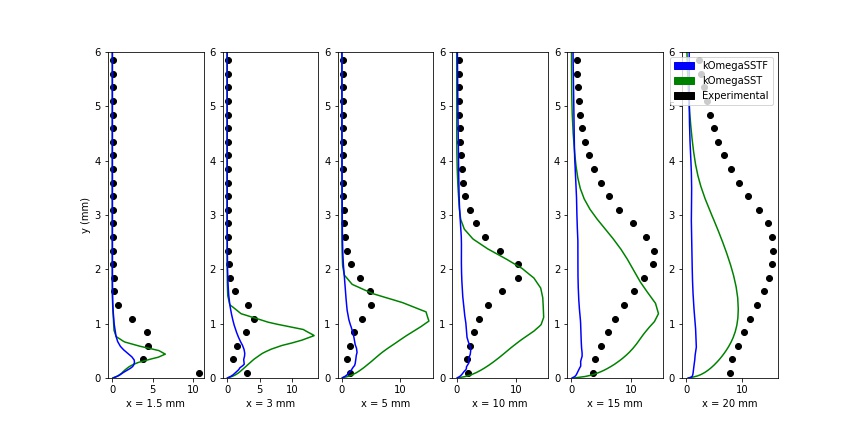}}}
\caption{Local comparisons between the experiments, standard k-$\omega$ SST calculation and k-$\omega$ SST calculation with the Reboud correction on the fine mesh for a) Reynolds shear stress and b) Turbulent Kinetic energy} \label{F_Tau12_SST}
\end{figure}

A local comparison is carried out for time-averaged velocities in the stream wise and the wall directions, the turbulent kinetic energy and the Reynolds shear stress at profiles as shown in Fig \ref{profiles}. The performance of both models to predict the time-averaged velocity in the stream wise direction agrees well with experiments while approaching downstream (see Fig \ref{UMeanX_SST} (a)). There are minor discrepancies close to throat where cavitation occurs.However, comparing the data for the wall directions in Fig. \ref{UMeanX_SST} (b), both the standard and corrected models initially co-relate with experiments but the prediction starts showing major discrepancies downstream, over-predicting the velocity. This is also reflected in local profiles for Reynolds wall shear stress where the fluctuations for velocity in wall direction are also accounted for. Figure \ref{Tau12_SST} (a) presents the comparisons for Reynolds shear stress. While the corrected model is slightly able to predict the wall shear stress, the standard k-$\omega$ SST model is unable to predict any such shear stress. A similar observation can be noted in plotting the local profiles for the turbulent kinetic energy (TKE) shown in Fig \ref{Tau12_SST} (b). While the corrected model is able to predict turbulent kinetic energy close to the throat where the primary cavity pinches off, the prediction shows large differences with experimental data downstream. On the other hand, the standard model displays no TKE. A significant change in observations from other similar studies \citep{zhang2021compressible} is noted as in the previous study, the corrected model displayed reduced Reynolds shear stress compared to the standard model.
To substantiate this conclusion, a similar comparison is carried for the k-$\omega$ SST SAS model with and without the Reboud correction on the same mesh and the  k-$\omega$ SST on the finer mesh .  A completely opposite observation is noted in Fig \ref{Tau12_SAS} (a) and (b). While the standard SAS model is able to predict the Reynolds shear stress close to the throat, the corrected model is not able to predict any shear stress. Likewise, it displays no TKE in Fig \ref{Tau12_SAS} (b) in contrast to the standard model which is able to predict the TKE, especially at the fourth position.  A similar observation is noted in Fig \ref{F_Tau12_SST} (a) where the standard model is able to predict the shear stress to some extent but overestimates it close to the throat while the corrected model is not able to predict any shear stress. Similarly, in Fig \ref{F_Tau12_SST} (b), the standard model is able to predict the TKE though with large discrepancies in the magnitude and position, the corrected model is not able to predict the TKE significantly. It can be posited that Reboud correction dampens the eddy viscosity and thus restrains the wall shear stress completely while the results from Fig \ref{Tau12_SST} (a) and (b) can be considered as anomalous.

\begin{figure}
\centering
{\resizebox*{15cm}{!}{\includegraphics{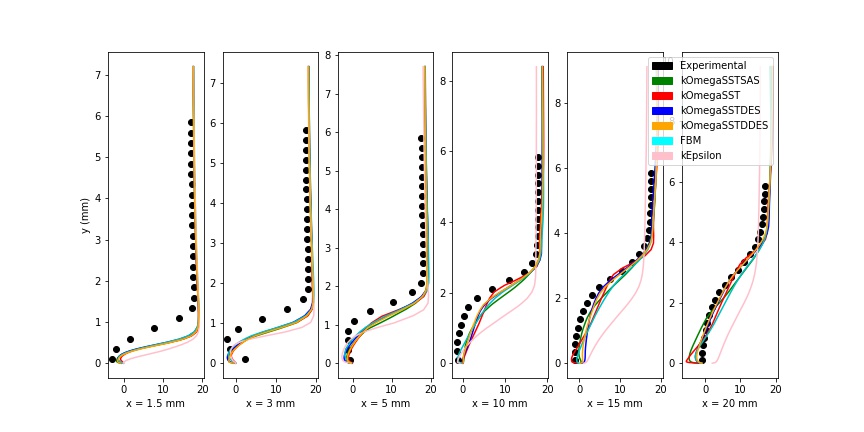}}}\hspace{8pt}
{\resizebox*{15cm}{!}{\includegraphics{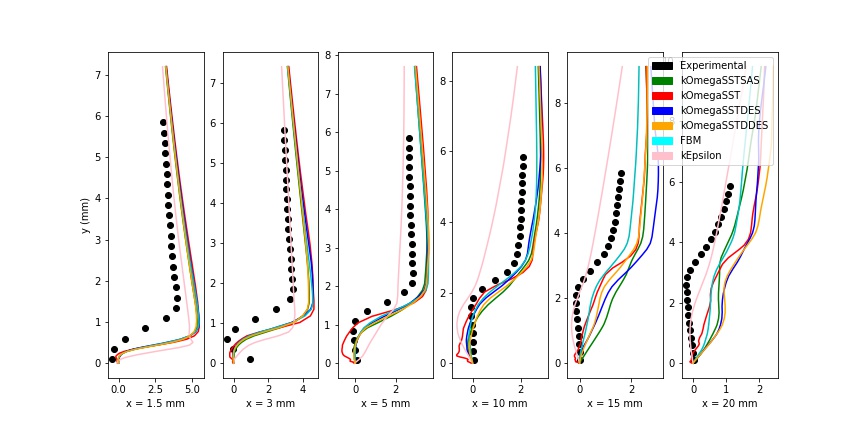}}}
\caption{Comparison for time-averaged velocity in both a) stream wise direction and b) wall direction} \label{2DI_UMeanX_total}
\end{figure}

\begin{figure}
\centering
{\resizebox*{15cm}{!}{\includegraphics{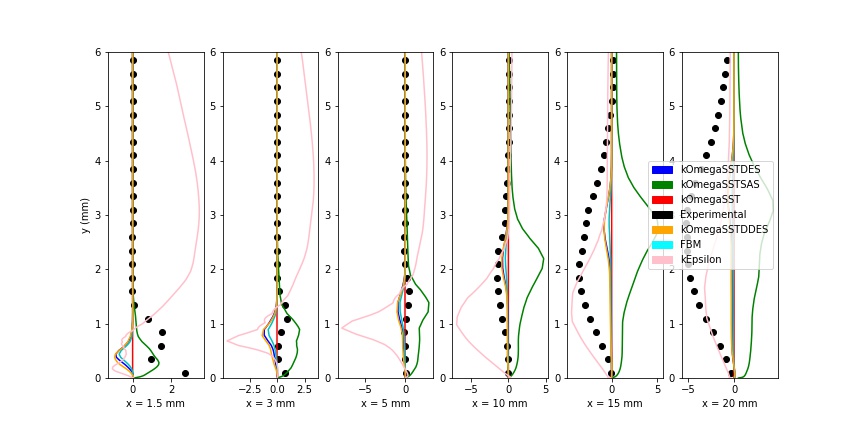}}}\hspace{8pt}
{\resizebox*{15cm}{!}{\includegraphics{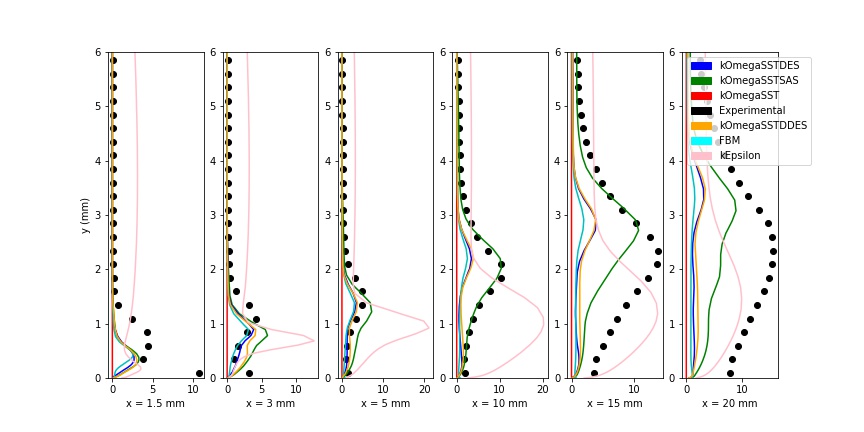}}}
\caption{Comparison for a) Reynolds shear stress and b) Turbulent kinetic energy for all models for intermediate mesh} \label{2DI_tau12_total}
\end{figure}

\begin{figure}
\centering
{\resizebox*{15cm}{!}{\includegraphics{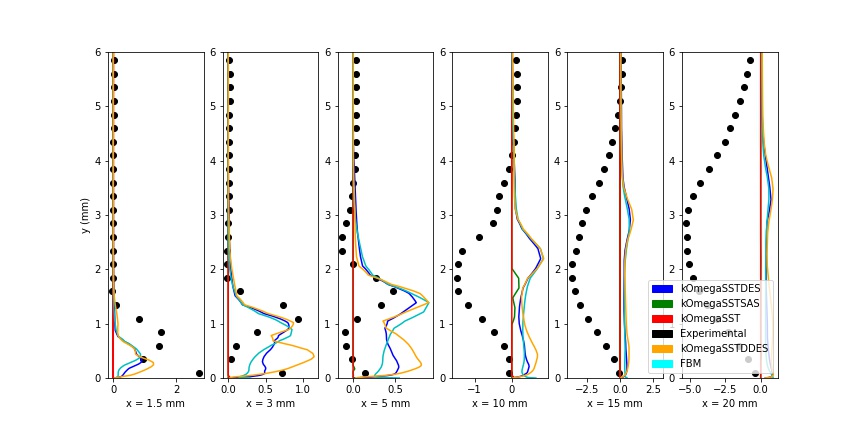}}}\hspace{8pt}
{\resizebox*{15cm}{!}{\includegraphics{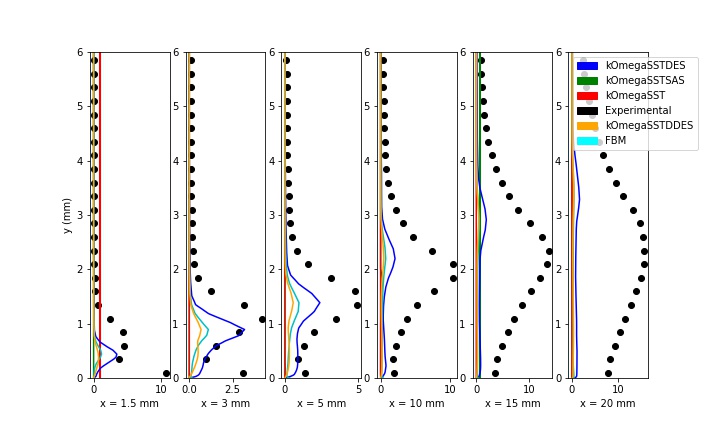}}}
\caption{Comparison for a) Reynolds shear stress and b) Turbulent kinetic energy for all models on the fine mesh} \label{2DF_Tau12_total}
\end{figure}

\begin{figure}
\centering
{\resizebox*{15cm}{!}{\includegraphics{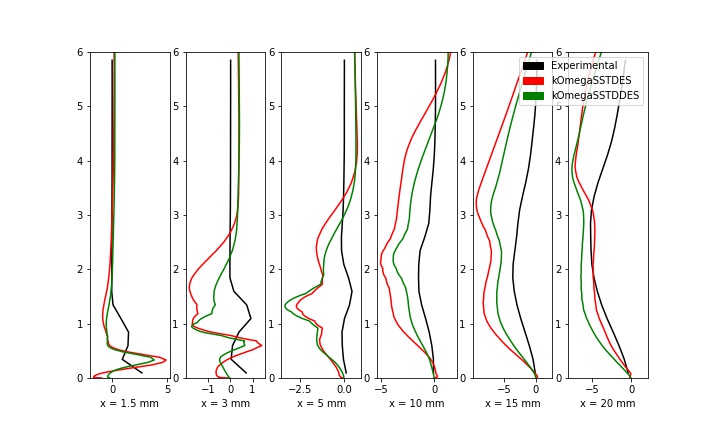}}}\hspace{8pt}
{\resizebox*{15cm}{!}{\includegraphics{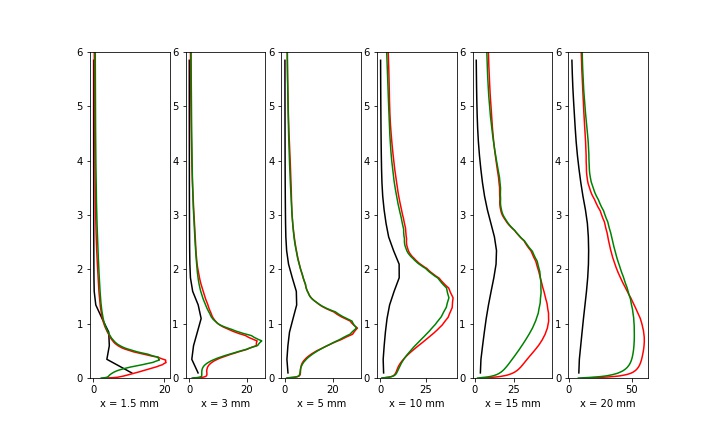}}}
\caption{Comparison for a) resolved Reynolds shear stress and b) resolved Turbulent kinetic energy for k-$\omega$ SSTDES and DDES models on the intermediate mesh } \label{2DI_Tau12_resolved}
\end{figure}

\begin{figure}
\centering
{\resizebox*{15cm}{!}{\includegraphics{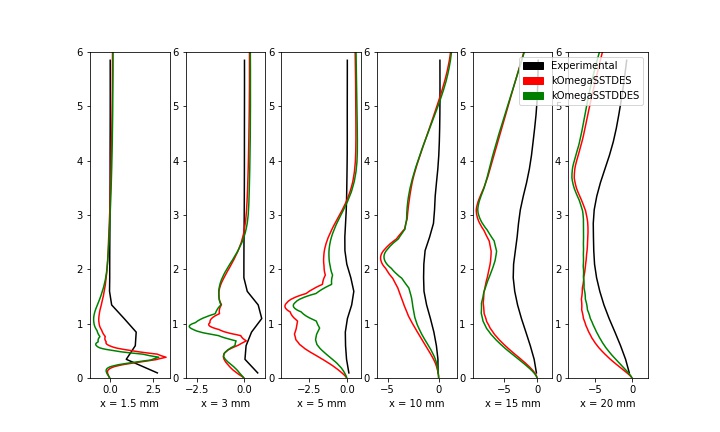}}}\hspace{8pt}
{\resizebox*{15cm}{!}{\includegraphics{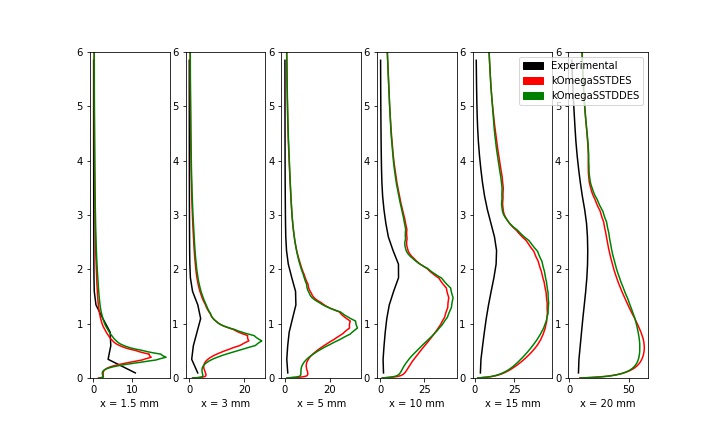}}}
\caption{Comparison for a) resolved Reynolds shear stress and b) resolved Turbulent kinetic energy for k-$\omega$ SSTDES and DDES models on the fine mesh} \label{2DF_Tau12_resolved}
\end{figure}

Comparing all the models for the local profiles, it can be remarked that all of them have near-identical behaviour. Figs \ref{2DI_UMeanX_total} (a) and (b) show the time-averaged velocity data from various models plotted against the experimental data.Regarding velocity in the streamwise direction, it is observed all models follow identical behaviour with the exception of the k-$\epsilon$ model that displays additional discrepancies downstream close to the wall. While the other models are able to reproduce the velocity curve similar to the experiments as it is measured farther from the throat and downstream, the k-$\epsilon$ model makes completely incorrect predictions downstream. Similarly in the wall direction, all models show identical behaviour, not able to predict accurately the experimental values with the sole anomaly being k-$\epsilon$ model that aligns the closest with experiments near the throat but significantly under-predicts it downstream. Similar observations can be noted for Reynolds shear stress and TKE plots in Fig \ref{2DI_tau12_total} (a) and (b). For the Reynolds shear stress, almost all the models display identical behaviour- while being able to predict the behaviour away from the wall, there are major differences between the experiments and models with the k-$\epsilon$ having significant differences. The TKE plots shows the models able to predict the pattern of TKE moving downstream and away from the wall albeit with lesser values than detected in experiments. Some of the simulations like the k-$\omega$ SST-SAS,  k-$\omega$ SST-DES and k-$\omega$ SST-DDES simulations are able to predict much improved TKE values than the other models. An important remark to be noted is despite the usage of hybrid models or finer grids (see Fig \ref{2DF_Tau12_total} (a) and (b)), standard models are unable to accurately predict turbulence properties as seen in experiments. The Reboud correction does seem to improve cavity shedding but upto an extent and at the cost of dampening the eddy viscosity and as a result, nullifying the Reynolds wall shear stress and TKE. 
Since k-$\omega$ SST-DES and k-$\omega$ SST-DDES models were also utilized in this study, it is inquisitive to see if the resolved parts of TKE and Reynolds shear stress for these models is able to replicate the experiment. Fig \ref{2DI_Tau12_resolved} (a) and (b) shows the resolved part of the Reynolds shear stress and turbulent kinetic energy respectively for both the models on the intermediate mesh. While the Reynolds shear stress predictions for both models show a very similar pattern though showing signs of difference downstream, both models are not able to predict the Reynolds shear stress produced in experiments. Comparing the TKE plots, both models do show a similar global trend of the TKE but over-estimate the magnitude. In the first subplot, the estimation is imprecise but is able to present the increase in kinetic energy at approximately the same position close to the throat where cavitation initiation commences. The TKE estimation works similarly at the other probe positions, being able to predict the position of the 'bump' having high TKE though with a much over-estimated position. This 'bump' corresponds to the cloud cavity formation near the throat, its expansion in size before shedding and finally collapsing downstream in an otherwise stable turbulent flow. However, the model predictions are completely askew in the final profile, 20 mm from the throat where they predict the cavity closer to the venturi wall rather than away from the wall.  Fig \ref{2DF_Tau12_resolved} (a) and (b) show a similar analysis for both the models on the finer mesh.  Both models are unable to predict the Reynolds shear stress and while they are able to predict the position of the cavity from the TKE plots, they overestimate the TKE values. An interesting point of discussion is the prediction by the turbulence models. The DDES model was originally proposed to ameliorate the grid-induced separation seen in the DES model. However, both models display analogous profiles and this behaviour is repeated across both mesh types. Another point of discussion arises with the region of formation of the cloud cavity. The cloud cavity forms, detaches, sheds and collapses very close to the venturi wall and throat- a region, a part of whose could be blended in the RANS-modelling region. Thus, some part of the cloud cavity region would be modelled in a RANS zone and the other part in a LES zone. This would directly influence the velocity in the wall direction and thus, directly the wall component of TKE and indirectly the Reynolds wall shear stress. Running the DES and DDES models on finer grids and then comparing the fluctuating components would be further conducted.

\section{Conclusion}\label{conclusion}

In this work, a comprehensive review of various standard RANS and hybrid RANS-LES models with and without compressibility effects to predict cloud cavitating flow inside a venturi channel by comparing the turbulent quantities  globally and locally with data from experiments was conducted. It was found the Reboud correction reduces the eddy viscosity over-prediction and ensure periodic vapor shedding but at finer meshes, the platform OpenFOAM's high dissipation rate reduces the eddy viscosity and confirms the unsteady vapor shedding for the standard models also. However, on a closer comparison of the turbulence quantities on a local scale, we observe huge discrepancies with data from experiments with the corrected models showing no fluctuations. This could be accounted to the fact that the Reboud correction dampens the eddy viscosity so such an extent where it makes the shear stress zero.  It is also notably interesting that while some simulations like PANS simulations at certain $f_{k}$ values were able to reproduce the cloud cavitating flow, some of them were unable to reproduce the cloud cavitating flow despite lowering the cavitation number $\sigma$. This sheds a little more light at the influence of the turbulence model on the cavitation dynamics.
The discrepancies could also be accounted to 3D effects, the choice of turbulence models and the choice of cavitation models. However, switching the turbulence models to hybrid ones improves the cavitation prediction in initial meshes, it is also observed that there is no major improvement after a certain level of refinement with the discrepancies in the turbulence properties still persisting. It can be concluded that hybrid models are able to improve the modelling up to a certain extent only.
There may exist a further analysis to refine the mesh more, in the case of DES models and for PANS make the filter ratio $f_{k}$ spatially varying as a factor of the turbulent length scale to improve the prediction. Another aspect of future scope for this study would be to use 3D grids. Though they currently possess a considerable computational cost challenge, techniques like adaptive mesh-refinement (AMR) could alleviate the obstacle. Regardless of all methods of mesh refinement and modelling setups, there seems to exist a bottleneck with both standard and hybrid models not able to accurately predict the cloud cavitation characteristics at the local scale. Data-driven techniques seem very promising to solve the above-stated bottleneck. The methods would utilise the experimental or DNS data as their input for their machine learning or deep learning algorithms and use the algorithm output for the traditional field input for turbulence models and are potentially promising. \citep{zhang2019recent, xu2021rans} 


\section{Appendix: Implementation of Turbulence models}
Many of the models discussed in this study are not part of the official OpenFOAM \citep{weller1998tensorial} code repository. Therefore, it was necessary to demonstrate that they have been implemented correctly , before running any study-related calculations. A simple case of the geometry used in the study is used for the verification. The flow inside the converging-diverging venturi is non-cavitating, yet turbulent. We conduct a k-$\omega$ SST calculation and a PANS calculation with $f_{k}$=1 calculation. Since $f_{k}$=1, signifies a URANS calculation, the results should be identical. Similar calculations have been conducted for the k-$\epsilon$ model and the Filter-based model (FBM). All the models have been run on the same mesh of 20,000 cells with a timestep size of 0.00001 for the same amount of computational time. To ensure no numerical scheme or solver affects the models, the schemes and solvers have been kept the same. 

\begin{figure}[H]
\centering
{\resizebox*{15cm}{!}{\includegraphics{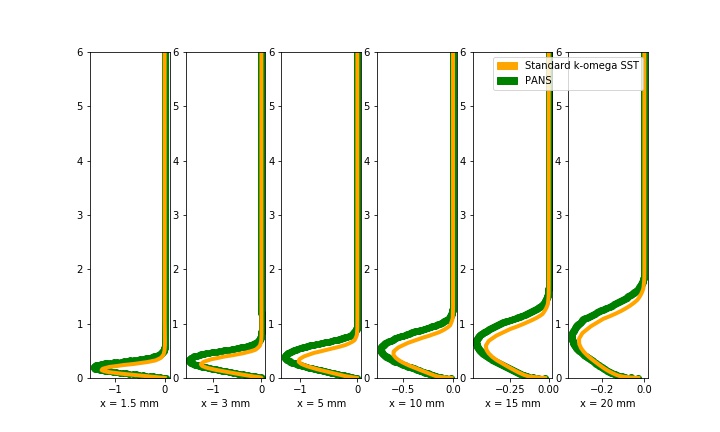}}}\hspace{8pt}
{\resizebox*{15cm}{!}{\includegraphics{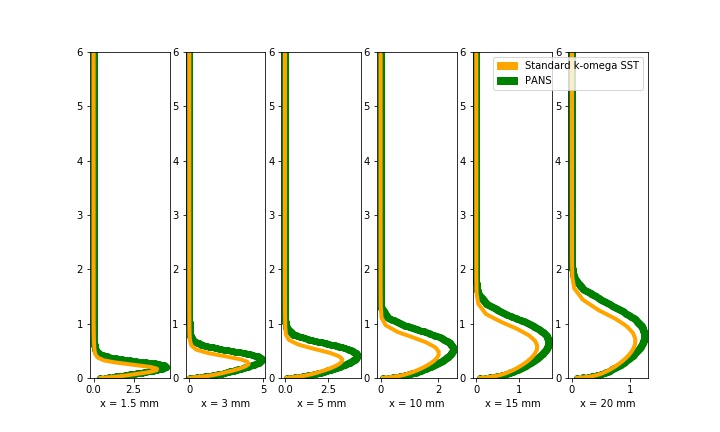}}}
\caption{Comparison for a) Reynolds shear stress and b) Turbulent kinetic energy for k-$\omega$ SST and PANS with $f_{k}$=1} \label{kSSTPANS}
\end{figure}

Figures \ref{kSSTPANS} (a) and (b) compare the Reynolds wall shear stress and the turbulent kinetic energy on a local scale for both the models. The data has been plotted using probes at 1.5 mm, 3mm, 5 mm, 10 mm, 15 mm and 20 mm respectively Both figures are identical thus validating that the models are implemented properly. A similar comparison is made for the Filter-Based Method (FBM) and the k-$\epsilon$ model on the same test case.

\begin{figure}[H]
\centering
{\resizebox*{15cm}{!}{\includegraphics{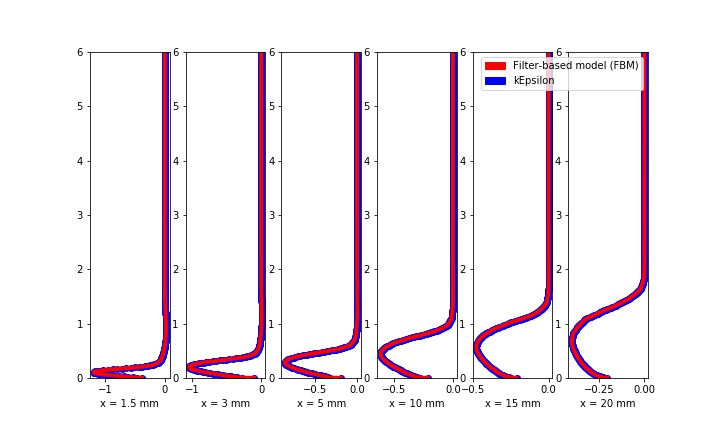}}}\hspace{8pt}
{\resizebox*{15cm}{!}{\includegraphics{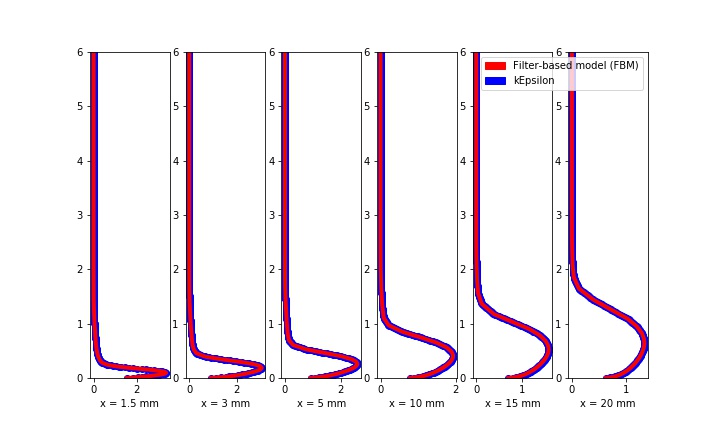}}}
\caption{Comparison for a) Reynolds shear stress and b) Turbulent kinetic energy for  a) FBM and b) k-$\epsilon$} \label{kfbm}
\end{figure}

Figures \ref{kfbm} (a) and (b) compare the Reynolds wall shear stress and the turbulent kinetic energy on a local scale for both the models. The data has been plotted on the same probes as stated above. Both sets of models look identical on the local level thus validating that the models have been implemented properly. 

\section{Funding}
This research did not receive any specific grant from funding agencies in the public, commercial, or not-for-profit sectors.

\section{Data Availability}

The data that support the findings of this study are available from the corresponding author upon reasonable request.

\section{Bibliography}
\bibliographystyle{elsarticle-num} 
\bibliography{references}

\begin{thebibliography}{10}
\expandafter\ifx\csname url\endcsname\relax
  \def\url#1{\texttt{#1}}\fi
\expandafter\ifx\csname urlprefix\endcsname\relax\def\urlprefix{URL }\fi
\expandafter\ifx\csname href\endcsname\relax
  \def\href#1#2{#2} \def\path#1{#1}\fi

\bibitem{zein2010modeling}
A.~Zein, M.~Hantke, G.~Warnecke, Modeling phase transition for compressible
  two-phase flows applied to metastable liquids, Journal of Computational
  Physics 229~(8) (2010) 2964--2998.

\bibitem{merkle1998computational}
C.~L. Merkle, Computational modelling of the dynamics of sheet cavitation, in:
  Proc. of the 3rd Int. Symp. on Cavitation, Grenoble, France, 1998, 1998.

\bibitem{kunz2000preconditioned}
R.~F. Kunz, D.~A. Boger, D.~R. Stinebring, T.~S. Chyczewski, J.~W. Lindau,
  H.~J. Gibeling, S.~Venkateswaran, T.~Govindan, A preconditioned
  {N}avier--{S}tokes method for two-phase flows with application to cavitation
  prediction, Computers \& Fluids 29~(8) (2000) 849--875.

\bibitem{sauer2000unsteady}
J.~Sauer, G.~H. Schnerr, Unsteady cavitating flow-a new cavitation model based
  on a modified front capturing method and bubble dynamics, in: Proceedings of
  2000 ASME Fluid Engineering Summer Conference, Vol. 251, 2000, pp.
  1073--1079.

\bibitem{gnanaskandan2016large}
A.~Gnanaskandan, K.~Mahesh, Large eddy simulation of the transition from sheet
  to cloud cavitation over a wedge, International Journal of Multiphase Flow 83
  (2016) 86--102.

\bibitem{bensow2010implicit}
R.~E. Bensow, G.~Bark, Implicit {LES} predictions of the cavitating flow on a
  propeller, Journal of Fluids Engineering 132~(4) (2010).

\bibitem{reboud1998two}
J.-L. Reboud, B.~Stutz, O.~Coutier, Two phase flow structure of cavitation:
  experiment and modeling of unsteady effects, in: 3rd International Symposium
  on Cavitation CAV1998, Grenoble, France, Vol.~26, 1998.

\bibitem{coutier2003evaluation}
O.~Coutier-Delgosha, R.~Fortes-Patella, J.-L. Reboud, Evaluation of the
  turbulence model influence on the numerical simulations of unsteady
  cavitation, Journal of Fluids Engineering 125~(1) (2003) 38--45.

\bibitem{coutier2003numerical}
O.~Coutier-Delgosha, J.~Reboud, Y.~Delannoy, Numerical simulation of the
  unsteady behaviour of cavitating flows, International Journal for Numerical
  Methods in Fluids 42~(5) (2003) 527--548.

\bibitem{seo2009numerical}
J.~Seo, S.~Lele, Numerical investigation of cloud cavitation and cavitation
  noise on a hydrofoil section, in: 7th International Symposium on Cavitation
  CAV2009, Ann Arbor, Michigan, USA, Vol.~2, 2009.

\bibitem{zhang2021compressible}
X.-L. Zhang, M.-M. Ge, G.-J. Zhang, O.~Coutier-Delgosha, Compressible effects
  modeling for turbulent cavitating flow in a small venturi channel: An
  empirical turbulent eddy viscosity correction, Physics of Fluids 33~(3)
  (2021) 035148.

\bibitem{hidalgo2019scale}
V.~Hidalgo, X.~Escaler, E.~Valencia, X.~Peng, J.~Erazo, D.~Puga, X.~Luo,
  Scale-adaptive simulation of unsteady cavitation around a {NACA}66 hydrofoil,
  Applied Sciences 9~(18) (2019) 3696.

\bibitem{chaouat2017state}
B.~Chaouat, The state of the art of hybrid rans/les modeling for the simulation
  of turbulent flows, Flow, turbulence and combustion 99~(2) (2017) 279--327.

\bibitem{heinz2020review}
S.~Heinz, A review of hybrid rans-les methods for turbulent flows: Concepts and
  applications, Progress in Aerospace Sciences 114 (2020) 100597.

\bibitem{wu2005time}
J.~Wu, G.~Wang, W.~Shyy, Time-dependent turbulent cavitating flow computations
  with interfacial transport and filter-based models, International Journal for
  Numerical Methods in Fluids 49~(7) (2005) 739--761.

\bibitem{zhang2016hybrid}
G.~Zhang, W.~Shi, D.~Zhang, C.~Wang, L.~Zhou, A hybrid {RANS/LES} model for
  simulating time-dependent cloud cavitating flow around a {NACA}66 hydrofoil,
  Science China Technological Sciences 59~(8) (2016) 1252--1264.

\bibitem{kim2009numerical}
S.-E. Kim, A numerical study of unsteady cavitation on a hydrofoil, in: 7th
  International Symposium on Cavitation CAV2009, Ann Arbor, Michigan, USA,
  Vol.~1, 2009.

\bibitem{bensow2011simulation}
R.~E. Bensow, Simulation of the unsteady cavitation on the {D}elft {T}wist11
  foil using {RANS,DES} and {LES}, in: Second International Symposium on Marine
  Propulsors, 2011.

\bibitem{vaz2017improved}
G.~Vaz, T.~Lloyd, A.~Gnanasundaram, Improved modelling of sheet cavitation
  dynamics on {D}elft {T}wist11 hydrofoil, in: MARINE VII: proceedings of the
  VII International Conference on Computational Methods in Marine Engineering,
  CIMNE, 2017, pp. 143--156.

\bibitem{huang2017numerical}
R.~Huang, X.~Luo, B.~Ji, Numerical simulation of the transient cavitating
  turbulent flows around the {C}lark-y hydrofoil using modified partially
  averaged {N}avier-{S}tokes method, Journal of Mechanical Science and
  Technology 31~(6) (2017) 2849--2859.

\bibitem{kanfoudi20183d}
H.~Kanfoudi, A.~Bel Hadj~Taher, R.~Zgolli, 3{D} analyze of the cavitation
  mechanism in turbulent flow using {P}artially-average {N}avier {S}tokes model
  around the {C}lark-y hydrofoil, Journal of Applied Fluid Mechanics 11~(6)
  (2018) 1637--1649.

\bibitem{ge2021cavitation}
M.~Ge, M.~Petkov{\v{s}}ek, G.~Zhang, D.~Jacobs, O.~Coutier-Delgosha, Cavitation
  dynamics and thermodynamic effects at elevated temperatures in a small
  venturi channel, International Journal of Heat and Mass Transfer 170 (2021)
  120970.

\bibitem{launder1983numerical}
B.~E. Launder, D.~B. Spalding, The numerical computation of turbulent flows,
  in: Numerical prediction of flow, heat transfer, turbulence and combustion,
  Elsevier, 1983, pp. 96--116.

\bibitem{wilcox1998turbulence}
D.~C. Wilcox, et~al., Turbulence modeling for CFD, Vol.~2, DCW industries La
  Canada, CA, 1998.

\bibitem{menter2003ten}
F.~R. Menter, M.~Kuntz, R.~Langtry, Ten years of industrial experience with the
  {SST} turbulence model, Turbulence, Heat and Mass transfer 4~(1) (2003)
  625--632.

\bibitem{spalart1997comments}
P.~R. Spalart, Comments on the feasibility of {LES} for wings, and on a hybrid
  {RANS/LES} approach, in: Proceedings of first AFOSR international conference
  on DNS/LES, Greyden Press, 1997.

\bibitem{spalart2006new}
P.~R. Spalart, S.~Deck, M.~L. Shur, K.~D. Squires, M.~K. Strelets, A.~Travin, A
  new version of detached-eddy simulation, resistant to ambiguous grid
  densities, Theoretical and Computational Fluid Dynamics 20~(3) (2006)
  181--195.

\bibitem{girimaji2003pans}
S.~S. Girimaji, R.~Srinivasan, E.~Jeong, P{ANS} turbulence model for seamless
  transition between {RANS} and {LES}: fixed-point analysis and preliminary
  results, in: Fluids Engineering Division Summer Meeting, Vol. 36975, 2003,
  pp. 1901--1909.

\bibitem{girimaji2006partially}
S.~S. Girimaji, Partially-averaged {N}avier-{S}tokes model for turbulence: A
  {R}eynolds-averaged {N}avier-{S}tokes to direct numerical simulation bridging
  method, Journal of Applied Mechanics 73 (2006) 413--431.

\bibitem{girimaji2006partially2}
S.~S. Girimaji, E.~Jeong, R.~Srinivasan, Partially averaged {N}avier-{S}tokes
  method for turbulence: {F}ixed point analysis and comparison with unsteady
  partially averaged {N}avier-{S}tokes, Journal of Applied Mechanics 73 (2006)
  422--429.

\bibitem{lakshmipathy2006partially}
S.~Lakshmipathy, S.~Girimaji, Partially-averaged {N}avier-{S}tokes method for
  turbulent flows: kw model implementation, in: 44th AIAA Aerospace Sciences
  Meeting and Exhibit, 2006, p. 119.

\bibitem{johansen2004filter}
S.~T. Johansen, J.~Wu, W.~Shyy, Filter-based unsteady {RANS} computations,
  International Journal of Heat and Fluid Flow 25~(1) (2004) 10--21.

\bibitem{menter2005scale}
F.~Menter, Y.~Egorov, A scale adaptive simulation model using two-equation
  models, in: 43rd AIAA Aerospace Sciences Meeting and Exhibit, 2005, p. 1095.

\bibitem{menter2010scale}
F.~Menter, Y.~Egorov, The scale-adaptive simulation method for unsteady
  turbulent flow predictions. part 1: theory and model description, Flow,
  Turbulence and Combustion 85~(1) (2010) 113--138.

\bibitem{kumar2020assessment}
A.~Kumar, A.~Ghobadian, J.~M. Nouri, Assessment of cavitation models for
  compressible flows inside a nozzle, Fluids 5~(3) (2020) 134.

\bibitem{leclercq2017numerical}
C.~Leclercq, A.~Archer, R.~Fortes-Patella, F.~Cerru, Numerical cavitation
  intensity on a hydrofoil for 3{D} homogeneous unsteady viscous flows,
  International Journal of Fluid Machinery and Systems 10~(3) (2017) 254--263.

\bibitem{li2019calculation}
X.~Li, B.~Li, B.~Yu, Y.~Ren, B.~Chen, Calculation of cavitation evolution and
  associated turbulent kinetic energy transport around a {NACA}66 hydrofoil,
  Journal of Mechanical Science and Technology 33~(3) (2019) 1231--1241.

\bibitem{geng2020assessment}
L.~Geng, X.~Escaler, Assessment of {RANS} turbulence models and {Z}wart
  cavitation model empirical coefficients for the simulation of unsteady cloud
  cavitation, Engineering Applications of Computational Fluid Mechanics 14~(1)
  (2020) 151--167.

\bibitem{chebli2021influence}
R.~Chebli, B.~Audebert, G.~Zhang, O.~Coutier-Delgosha, Influence of the
  turbulence modeling on the simulation of unsteady cavitating flows, Computers
  \& Fluids 221 (2021) 104898.

\bibitem{decaix2012time}
J.~Decaix, E.~Goncalves, Time-dependent simulation of cavitating flow with k- l
  turbulence models, International Journal for Numerical Methods in Fluids
  68~(8) (2012) 1053--1072.

\bibitem{decaix2013investigation}
J.~Decaix, E.~Goncalves, Investigation of three-dimensional effects on a
  cavitating venturi flow, International Journal of Heat and Fluid Flow 44
  (2013) 576--595.

\bibitem{ennouri2019numerical}
M.~Ennouri, H.~Kanfoudi, A.~Bel Hadj~Taher, R.~Zgolli, Numerical flow
  simulation and cavitation prediction in a centrifugal pump using an {SST-SAS}
  turbulence model, Journal of Applied Fluid Mechanics 12~(1) (2019) 25--39.

\bibitem{sedlar2016numerical}
M.~Sedlar, B.~Ji, T.~Kratky, T.~Rebok, R.~Huzl{\'\i}k, Numerical and
  experimental investigation of three-dimensional cavitating flow around the
  straight {NACA}2412 hydrofoil, Ocean Engineering 123 (2016) 357--382.

\bibitem{gao2012hybrid}
G.~H. Gao, J.~Zhao, F.~Ma, W.~D. Luo, Hybrid {RANS--LES} modeling for unsteady
  cavitating flow simulation, in: Applied Mechanics and Materials, Vol. 152,
  Trans Tech Publ, 2012, pp. 1187--1190.

\bibitem{wang2021comparative}
Z.~Wang, X.~Zhang, Y.~Wang, J.~Liu, Comparative study between turbulence models
  in unsteady cavitating flow with special emphasis on shock wave propagation,
  Ocean Engineering 240 (2021) 109988.

\bibitem{hwang2021numerical}
H.-S. Hwang, K.-J. Paik, S.-H. Lee, G.~Song, Numerical study on the vibration
  and noise characteristics of a {D}elft {T}wist11 hydrofoil, Journal of Marine
  Science and Engineering 9~(2) (2021) 144.

\bibitem{liu2021numerical}
J.~Liu, J.~Yu, Z.~Yang, Z.~He, K.~Yuan, Y.~Guo, Y.~Li, Numerical investigation
  of shedding dynamics of cloud cavitation around 3{D} hydrofoil using
  different turbulence models, European Journal of Mechanics-B/Fluids 85 (2021)
  232--244.

\bibitem{sun2019numerical}
T.~Sun, X.~Zhang, C.~Xu, G.~Zhang, S.~Jiang, Z.~Zong, Numerical modeling and
  simulation of the shedding mechanism and vortex structures at the development
  stage of ventilated partial cavitating flows, European Journal of
  Mechanics-B/Fluids 76 (2019) 223--232.

\bibitem{ji2013numerical}
B.~Ji, X.~Luo, Y.~Wu, X.~Peng, Y.~Duan, Numerical analysis of unsteady
  cavitating turbulent flow and shedding horse-shoe vortex structure around a
  twisted hydrofoil, International Journal of Multiphase Flow 51 (2013) 33--43.

\bibitem{zhou2019development}
H.~Zhou, M.~Xiang, S.~Zhao, W.~Zhang, Development of a multiphase cavitation
  solver and its application for ventilated cavitating flows with natural
  cavitation, International Journal of Multiphase Flow 115 (2019) 62--74.

\bibitem{goncalves2011numerical}
E.~Goncalv{\`e}s, Numerical study of unsteady turbulent cavitating flows,
  European Journal of Mechanics-B/Fluids 30~(1) (2011) 26--40.

\bibitem{weller1998tensorial}
H.~G. Weller, G.~Tabor, H.~Jasak, C.~Fureby, A tensorial approach to
  computational continuum mechanics using object-oriented techniques, Computers
  in Physics 12~(6) (1998) 620--631.

\bibitem{girimaji2005partially}
S.~Girimaji, K.~Abdol-Hamid, Partially-averaged {N}avier {S}tokes model for
  turbulence: {I}mplementation and validation, in: 43rd AIAA Aerospace Sciences
  Meeting and Exhibit, 2005, p. 502.

\bibitem{zhang2019recent}
X.~Zhang, J.~Wu, O.~Coutier-Delgosha, H.~Xiao, Recent progress in augmenting
  turbulence models with physics-informed machine learning, Journal of
  Hydrodynamics 31~(6) (2019) 1153--1158.

\bibitem{xu2021rans}
M.~Xu, H.~Cheng, B.~Ji, R{ANS} simulation of unsteady cavitation around a
  {C}lark-{Y} hydrofoil with the assistance of machine learning, Ocean
  Engineering 231 (2021) 109058.

\end{thebibliography}
\end{document}